\newlength{\intwidth}

\documentclass{jfm}
\usepackage{amssymb}

\usepackage{array}
\usepackage{amsmath}
\usepackage{latexsym}
\usepackage{bezier}
\usepackage{psfrag,graphicx}
\usepackage{bm}
\usepackage{float}
\usepackage{multirow}
\usepackage{mathrsfs}
\usepackage{color}

\begin{document}

\title[High-speed shear driven dynamos. Part 1. Asymptotic analysis]{High-speed shear driven dynamos. Part 1. Asymptotic analysis}
\author{Kengo Deguchi}
\affiliation{
School of Mathematical Sciences, Monash University, VIC 3800, Australia
}

\maketitle

\begin{abstract}
Rational large Reynolds number matched asymptotic \textcolor{black}{expansions} of three-dimensional \textcolor{black}{nonlinear magneto-hydrodynamic (MHD) states} are concerned.
The \textcolor{black}{nonlinear MHD states}, assumed to be predominantly driven by a unidirectional \textcolor{black}{shear,} can be sustained without any linear instability of the base flow \textcolor{black}{and hence are responsible for subcritical transition to turbulence}. 
Two classes of \textcolor{black}{nonlinear MHD states} are found. 
The first class of \textcolor{black}{nonlinear states} emerged out of a nice combination of the purely hydrodynamic vortex/wave interaction theory by Hall \& Smith (1991) and the resonant absorption theories \textcolor{black}{on} Alfv\'en waves, developed in the solar physics community \textcolor{black}{(e.g. Sakurai et al. 1991; Goossens et al. 1995).}
Similar to the hydrodynamic theory, the mechanism of the \textcolor{black}{MHD states} can be explained by the successive interaction of the roll, streak, and wave fields, which are now defined both for \textcolor{black}{the} hydrodynamic and magnetic fields. 
The derivation of this `vortex/Alfv\'en wave interaction' state is rather straightforward as the scalings for both of the hydrodynamic and magnetic fields are identical. 
It turns out that the leading order magnetic field of the asymptotic states appears only when a small external magnetic field \textcolor{black}{is present.}
However, it does not mean \textcolor{black}{that} purely \textcolor{black}{shear-driven} dynamos are not possible.
In fact, the second class of  `self-sustained shear driven dynamo theory' shows the magnetic generation \textcolor{black}{that is} slightly smaller size in the absence of any external field.
\textcolor{black}{Despite small size, the magnetic field} causes the novel feedback mechanism \textcolor{black}{in} the velocity field through resonant absorption, \textcolor{black}{wherein} the magnetic wave becomes more strongly amplified than the hydrodynamic counterpart.
\end{abstract}

\section{Introduction}

We are concerned with \textcolor{black}{the} mathematical descriptions of high-speed nonlinear magneto-hydrodynamic (MHD) flows driven by simple subcritical shear flows such as \textcolor{black}{the} plane Couette flow. 
MHD turbulence states occur in electrically conductive fluid flows (e.g. plasma or liquid metal) and are made of various complicated vortex structures. The interaction \textcolor{black}{between} the velocity and magnetic fields in the flow has attracted much attention as it \textcolor{black}{can} produce Ôdynamo statesÕ \textcolor{black}{that}
serve \textcolor{black}{as} an efficient way to convert kinetic energy to \textcolor{black}{magnetic energy.} 
The dynamo states are of central importance in \textcolor{black}{geophysics} and in astrophysics where such energy conversion \textcolor{black}{mechanisms are} common.
However, most of the mathematical dynamo studies are limited to the kinematic situation where the magnetic field is so small that it does not affect the hydrodynamic field. 

The strong nonlinear coupling of the velocity and magnetic fields in the MHD turbulence implies that we must greatly rely on numerical analysis, from which it is difficult to infer general properties.
The numerical dynamo studies aiming to compute the saturated magnetic field from the kinematic dynamos use meticulously designed complicated flow geometries in order to avoid various anti-dynamo theorems (see the review by Brandenburg \& Subramanian (2005) and references therein) or, otherwise, assume some linear instability of the flow (e.g. magneto-convection).
Numerical studies of dynamos in linearly stable flows are rather rare (Rincon et al. 2007; Rincon et al. 2008; Riols et al. 2013; Neumann \& Blackman 2017).

Another difficulty of fully computational study lies in the fact that the high Reynolds number of practical relevance hinders the simulation that requires very fine grid points. In order to analyse high Reynolds number MHD turbulence locally, a number of isotropic homogeneous simulations are performed with some artificial external forcing pouring energy into the hydrodynamic field (see Kida et al. 1991, Brandenburg et al. 2012, for example). 
However, it is not clear how those local MHD turbulent states are sustained in physically realisable flow configurations. 
\textcolor{black}{Though} modelling approaches are used to simplify or modify the governing equations to perform global high Reynolds number simulations, \textcolor{black}{the} introduction of some artificial assumptions \textcolor{black}{becomes} inevitable \textcolor{black}{in order to close the model system} (e.g. large eddy simulations, turbulent models, mean-field models).

To make some theoretical progresses towards the rational high-speed dynamo theory using minimal assumptions, in this paper, we employ the large Reynolds number matched asymptotic expansion of viscous-resistive MHD equations. 
The large hydrodynamic and magnetic Reynolds numbers limit is known as the singular limit of the governing equations, and so, the leading-order solution may possess singularities. 
The small-scale flow induced near the singular point must behave \textcolor{black}{in} a consistent manner as the large-scale flows, so this is exactly the place where we need the matching.
That matching condition is the crux to close the asymptotically reduced system, which has no need for artificial assumption or tuning parameters unlike heuristic model approaches.
In fact, the method on which our analysis is based is the cornerstone of progress \textcolor{black}{in} understanding high Reynolds number solutions of the Navier-Stokes and resistive MHD equations although, nowadays, only a few researchers are working in this field due to \textcolor{black}{the} technical complexity associated with the singularity. 
In the purely hydrodynamic study, such \textcolor{black}{a} singular point is known as \textcolor{black}{the} `critical layer', which occurs whenever the speed of the background flow coincides with the phase speed of the instability wave.
For the simplest linear case, the delicate interaction of the small- and large-scale flows through the matching conditions was established \textcolor{black}{for the first time} by Lin (1945, 1955).

In \textcolor{black}{the} solar physics community, the similar singular behaviour of wave-like instabilities has been studied \textcolor{black}{under} the resonant absorption theory. 
Since Alfv\'en (1942), a number of researchers studied \textcolor{black}{the} instability waves riding on uniform background velocity and magnetic fields. 
However, \textcolor{black}{the} solar atmosphere is highly non-uniform in reality, so the uniform flow assumption used in Alfv\'en (1942) is actually not valid.
The consideration of non-uniform background magnetic fields led to the finding of the Alfv\'en resonant point, where the wave behaves \textcolor{black}{in} a singular manner at the limit of large hydrodynamic and magnetic Reynolds numbers. 
The asymptotic structure of \textcolor{black}{the} linear disturbances excited by a background magnetic field was derived \textcolor{black}{by} Sakurai et al. (1991) \textcolor{black}{and} Goossens et al. (1995).
The theory and its extended results to steady non-zero background shear flows (Goossens et al. (1992), Erd\'elyi et al. (1995), Erd\'elyi (1997)) successfully describe the inner structure of the thin dissipative layer surrounding the resonant point, and how it should be matched to the outer Alfv\'en wave solution. 
Despite the similarity, the nature of the singularities was found to be mathematically different \textcolor{black}{from} the purely hydrodynamic critical layer.

\textcolor{black}{
To date, the majority of the existing large Reynolds number nonlinear hydrodynamic/MHD asymptotic theories 
assume absence of $O(1)$ feedback effect from the instability wave to the background flow.
That assumption is typical when 
the background flow varies only in one direction and, hence, the wave-like perturbation is two-dimensional; see Smith \& Bodonyi (1982), Clack \& Ballai (2009), Goossens \textcolor{black}{et al.} (2011), Deguchi \& Walton (2018).
For those asymptotic studies the interaction occurs merely one way from the background flow to the induced wave, although, in turbulent flows of course, the background flow profile can be largely distorted to cause some feedback effect on the wave instability. 
For two dimensional flows the feedback effect is difficult to implement theoretically, as too strong wave amplitude destroys the basic assumption that the critical or resonant layers are sufficiently thin. }
\begin{figure}
\centering
\includegraphics[scale=0.4,clip]{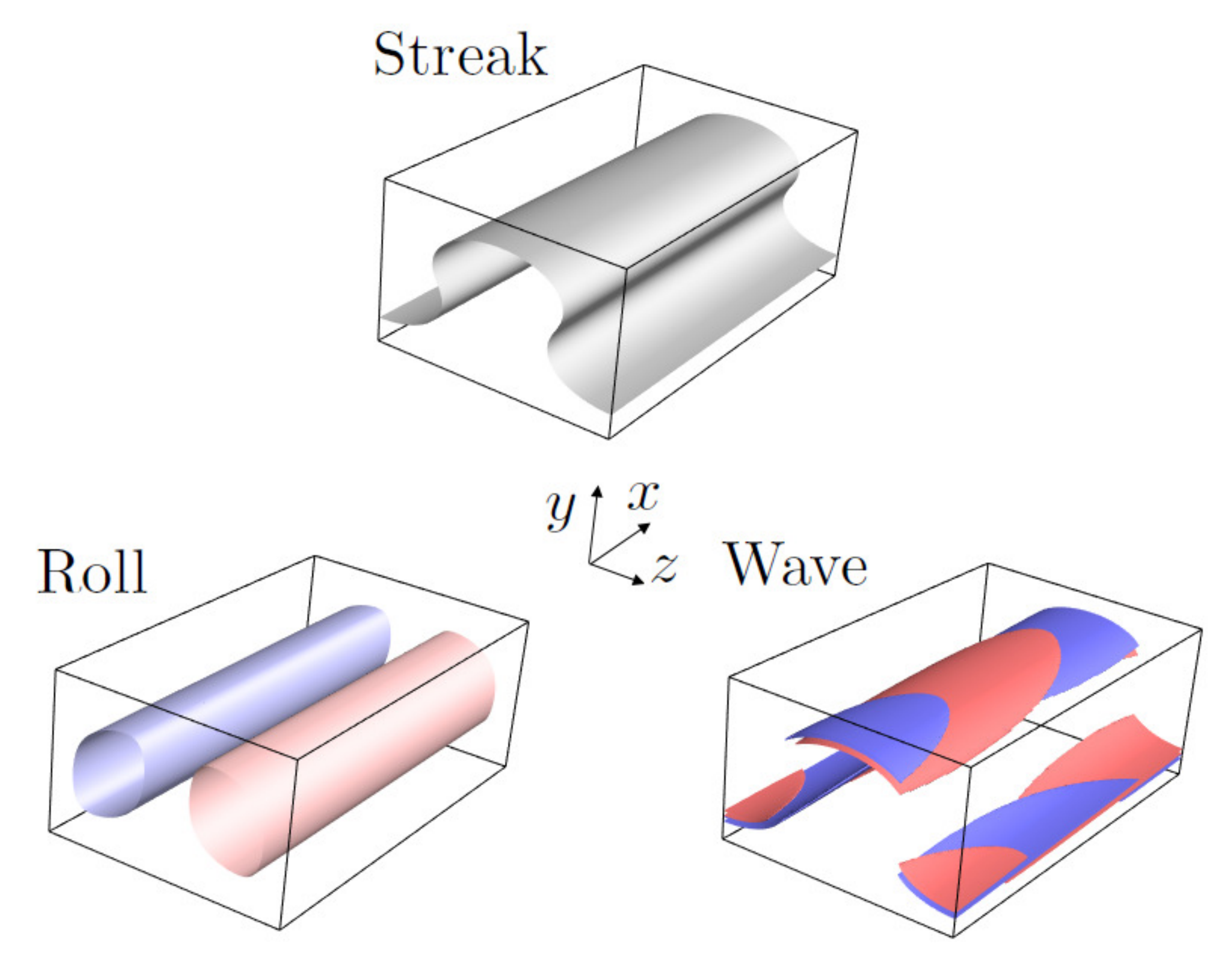}
\label{fig:bif}
\caption{The nonlinear steady solution of plane Couette flow at Reynolds number $16000$ (streamwise wavenumber 1.2, spanwise wavenumber $2$, upper branch; see  Deguchi \& Hall (2014) for definition). Here $x$ is taken to be the direction of the base flow; see section 2.
The wave field is visualised by 30\% maximum absolute streamwise vorticity (red is positive, blue is negative).
The roll-streak field corresponds to the streamwise average of the total flow field. 
The streak field is the zero locus of the streamwise velocity, whilst the roll field is the 
isosurfaces of the stream function for the cross-stream components of the averaged velocity (red is positive, blue is negative).
}
\end{figure}

It has been pointed out by \textcolor{black}{Hall, Smith and their colleagues between the} late 80's \textcolor{black}{and} early 90's that the three-dimensionality of the flow allows us to incorporate that feedback effect to the theory without destroying the critical layer structure.
Their theory, called the vortex/wave interaction theory \textcolor{black}{(Hall \& Smith 1990; Bennett, Hall \& Smith 1991)}, was originally found for boundary layer flows by considering the mutual interaction between the streamwise vortices developing slowly and \textcolor{black}{the} much rapid three-dimensional viscous (Tollmien-Shlichting) waves; see \textcolor{black}{Smith (1979), Hall (1983) and Hall \& Smith (1988)} also.
Subsequently, those authors noticed that similar interaction is possible for inviscid (Rayleigh) waves as well (Hall \& Smith 1991). 
What is remarkable in the latter type of the theory is that it rationally describes the mechanism by which the small wave significantly modifies the background flow to leading order.
This means that the large Reynolds number approximation has a potential to interpret the mysterious mechanism of nonlinear subcritical instability that can destabilise the linearly stable shear flows, \textcolor{black}{for example the} plane Couette flow. 
It was Waleffe (1997) who firstly recognised that possibility, although what he independently proposed was rather \textcolor{black}{a} heuristic model approach \textcolor{black}{that he termed} `self-sustaining process'.
His theory has rapidly \textcolor{black}{spread} among fluid dynamisists as it concisely explains the sustaining mechanism of the nonlinear states \textcolor{black}{through} the successive interaction of the roll, streak, and wave. Here, the streak and \textcolor{black}{the} roll constitutes the streamwise and the cross-stream components of the vortex component in the vortex/wave interaction; see figure 1, where the roll, streak, and wave components \textcolor{black}{as} computed in Deguchi \& Hall (2014).
Despite the great success of the self-sustaining process, an important ingredient \textcolor{black}{had been} overlooked until Wang, Gibson \& Waleffe (2007) - the critical layer structure. 
Subsequently, Hall \& Sherwin (2010) showed that the self-sustaining process is actually equivalent to mathematically rational \textcolor{black}{frameworks} of vortex/wave interaction if the former theory is supplemented by the critical layer analysis.
Two independently developed research streams over the last two decades, both of which \textcolor{black}{are} greatly inspired by earlier Benney (1984) and Benney \& Chow (1989), have now been integrated together.

The primal aim of this paper is to develop a nonlinear three-dimensional asymptotic \textcolor{black}{MHD} theory by replacing the Rayleigh wave in \textcolor{black}{the Hall--Smith theory} by the Alfv\'en wave. 
The formulation of this `vortex/Alfv\'en wave interaction theory' would be rather straightforward, 
\textcolor{black}{as} the particular form of the viscous resistive MHD equations suggests that we should be able to apply the roll-streak-wave decomposition to the magnetic fields as well as the velocity fields.
Indeed, Rincon et al. (2007), Rincon et al. (2008), Riols et al. (2013) \textcolor{black}{have pointed out} some similarities of the nonlinear dynamo theory and the self-sustaining process
(However, it should be noted that those works treat dynamos driven by the magneto-rotational instability, while the effect of \textcolor{black}{the} system rotation is not considered here. We shall comment on \textcolor{black}{the} implication of our work on the rotational case in section 5).

After formulating our problem in the next section, we begin the asymptotic analysis in section 3, assuming the largeness of the Reynolds number. 
We shall see that the Alfv\'en wave produced in inhomogeneous mean fields, namely hydrodynamic and magnetic streaks, can be analysed using the resonant absorption theory. 
Interestingly, the size of the wave in general differs from the purely hydrodynamic case \textcolor{black}{as} the interaction \textcolor{black}{between} the wave \textcolor{black}{and} the vortex now occurs through the Alfv\'en resonant layer rather than the critical layer. 
In the same section, the sustained mechanism of the \textcolor{black}{vortex/Alfv\'en wave interaction states} will be examined to show that unless \textcolor{black}{a} small magnetic field is applied, the vortex/Alfv\'en wave interaction state is not possible. 
However, \textcolor{black}{this} does not mean that the dynamo states cannot be realised in the zero external magnetic field limit. 
In fact, the numerical study of the shear driven dynamos in the companion paper \textcolor{black}{Deguchi (2019)} showed the production of the magnetic field at that limit.
Motivated by that numerical result, section 4 \textcolor{black}{is concerned with} the mathematical description of such self-sustained shear driven dynamos, and show that the induced magnetic field is indeed slightly smaller than the vortex/Alfv\'en wave interaction states.
Finally, in section 5, we draw some conclusions.

\section{Formulation of the problem}

Throughout the paper, we use the Cartesian coordinates $(x_*,y_*,z_*)$. 
Consider incompressible viscous resistive MHD equations for the velocity field $\mathbf{v}_*$, the magnetic field $\mathbf{b}_*$, the current density $\mathbf{j}_*$, and the pressure $p_*$:
\begin{subequations}
\begin{eqnarray}
(\partial_{t_*}+\mathbf{v}_*\cdot \nabla_*)\mathbf{v}_*
=\frac{1}{\rho}(-\nabla_* p_* +\mathbf{j}_*\times \mathbf{b}_*)+ \nu \nabla_*^2 \mathbf{v}_* ,\\
\partial_{t_*}\mathbf{b}_* 
= \nabla_* \times ( \mathbf{v}_*\times \mathbf{b}_*- \eta \mathbf{j}_*),\\
\nabla_* \cdot \mathbf{v}_*=0,\\
\mathbf{j}_*=\frac{1}{\mu_0}(\nabla_* \times \mathbf{b}_*),
\end{eqnarray}
\end{subequations}
where $\nabla_*=(\partial_{x_*},\partial_{y_*},\partial_{z_*})$, $\rho$ is the fluid density , $\nu$ is the fluid kinematic viscosity, $\eta$ is the fluid \textcolor{black}{electrical resistivity}, and $\mu_0$ is the fluid magnetic permeability. 

We assume that the flow is predominantly driven by some unidirectional hydrodynamic base flow, and take $U_0, L_0$ to be the typical \textcolor{black}{velocity magnitude} and the spatial scale of it.  Using the length scale $L_0$ and the velocity scale $U_0$, we define the non-dimensional variables as
\begin{subequations}
\begin{eqnarray}
\mathbf{v}_*=U_0 \mathbf{v}, \qquad \mathbf{b}_*=\sqrt{\mu_0 \rho}U_0\mathbf{b}, \qquad p_*=\rho U_0^2 p,\\
(x_*,y_*,z_*)=L_0(x,y,z),\qquad t_*=(L_0/U_0)t,
\end{eqnarray}
\end{subequations}
to get the non-dimensional version of the governing equations
\begin{subequations}\label{govcomp}
\begin{eqnarray}
\frac{D\mathbf{v}}{Dt}
- (\mathbf{b}\cdot \nabla)\mathbf{b}
&=&-\nabla q + \frac{1}{R} \nabla^2 \mathbf{v} ,\\
\frac{D\mathbf{b}}{Dt}-(\mathbf{b}\cdot \nabla)\mathbf{v}
&=&\frac{1}{R_m} \nabla^2 \mathbf{b}\label{ind1},\\
 \nabla \cdot  \mathbf{v}&=&0,
\end{eqnarray}
where $\nabla=(\partial_{x},\partial_{y},\partial_{z})$, $D/Dt=(\partial_t+\mathbf{v}\cdot \nabla)$ and
\begin{eqnarray}
 q=p+\frac{|\mathbf{b}|^2}{2}\label{totalpressure}
\end{eqnarray}
\end{subequations}
is the total pressure.
From the induction equation (\ref{ind1}), we can show that if the solenoidal condition $\nabla \cdot \mathbf{b}=0$ holds at \textcolor{black}{a} certain instant, then it should be always satisfied for all $t$. 

The flow is governed by the hydrodynamic Reynolds number $R$ and the magnetic Reynolds number $R_m$
\begin{eqnarray}
R=\frac{U_0 L_0 \rho_0}{\mu},\qquad R_m=\frac{U_0L_0 \mu_0}{\eta}.
\end{eqnarray}
The ratio of those parameters is known as the magnetic Prandtl number 
\begin{eqnarray}
P_m=R_m/R.
\end{eqnarray}

The theory to be developed can be applied for quite \textcolor{black}{a} wide range of shear flows but here, for the sake of simplicity and definiteness, we consider \textcolor{black}{the} plane Couette flow. 
Following the convention of the shear flow study, we take $x,y,z$ directions to be streamwise, wall-normal, and spanwise directions, respectively, assuming the periodicity of the flow in $x,z$ and the presence of the no-slip and perfectly insulating walls at $y=\pm 1$. The system has a laminar base flow solution $\mathbf{v}=U_b(y)\mathbf{e}_x$ with $U_b=y$, while our main interest is other nonlinear solutions.

If the no-slip conditions are replaced by certain generalised periodic conditions, the system becomes \textcolor{black}{a} `shearing box' frequently used in the local analyses of \textcolor{black}{hydrodynamical and astrophysical flows (Hawley et al. 1995; Brandenburg et al. 1995; Pumir 1996; Rincon et al. 2008; Riols et al. 2013; Sekimoto et al. 2016). 
The system admits time periodic solutions that tend to travelling waves as the Reynolds number gets large. Hall (2018) recently extended the vortex wave interaction theory to represent a family of vertically periodic vortex arrays in turbulent flows, and it is likely that the theory can be used to describe the limiting shearing box solution. 
In order to extend the theory modification is only necessary for the boundary conditions of the asymptotic system.
 The similar extension in the MHD version is fairly easy as we shall see later.
}
Note that our notation is different from the common notation used in the astrophysics community, where the azimuthal (streamwise), radial (vertical), and axial (spanwise) directions of the computational box are denoted by $y,x,z$, respectively (e.g. Riols et al. 2013).

\section{The vortex/Alfv\'en wave interaction theory}

Now let us construct the large $R$ asymptotic theory for the nonlinear three-dimensional MHD \textcolor{black}{states}. 
We assume that the nonlinear state is a travelling wave propagating in $x$ with a phase speed $s$. 
Here and hereafter, we apply the Galilean transform to convert the travelling wave to the steady state. 
Redefining the coordinate $x$, the governing equations (\ref{govcomp}) can be written in component form
\begin{subequations}\label{incompgov}
\begin{eqnarray}
~[(\mathbf{v}-s\mathbf{e}_x)\cdot \nabla]
\left[ \begin{array}{c} u\\ v\\ w \end{array} \right]
-[\mathbf{b}\cdot \nabla]
\left[ \begin{array}{c} a\\ b\\ c \end{array} \right]
=
-\left[ \begin{array}{c} q_x\\ q_y\\ q_z \end{array} \right]
+R^{-1} \nabla^2 \left[ \begin{array}{c} u\\ v\\ w \end{array} \right] \label{govmomentum}
\\
~[(\mathbf{v}-s\mathbf{e}_x)\cdot \nabla]
\left[ \begin{array}{c} a\\ b\\ c \end{array} \right]
-[\mathbf{b}\cdot \nabla]
\left[ \begin{array}{c} u\\ v\\ w \end{array} \right]=
R_m^{-1} \nabla^2 \left[ \begin{array}{c} a\\ b\\ c \end{array} \right] \\
u_x+v_y+w_z=0,\qquad a_x+b_y+c_z=0,
\end{eqnarray}
\end{subequations}
where $\mathbf{v}=[u,v,w], \mathbf{b}=[a,b,c]$.
In the following large $R$ asymptotic analysis, we assume that $P_m\sim O(1)$ or larger. 
The analyses for small $P_m$ cases are given in Appendix B
\textcolor{black}{(valid when $P_m$ is $O(R^{-1})$ or smaller, but needs a very large external magnetic field to maintain the three-dimensional magnetic field).}

The use of analogy from the Hall-Smith theory yields 
the asymptotic expansion 
\begin{eqnarray}
\left[ \begin{array}{c} u\\ v\\ w\\a\\b\\c\\q \end{array} \right]=
\left[ \begin{array}{c} U_b+P_m^{-1}\overline{u}\\ R_m^{-1}\overline{v}\\ R_m^{-1}\overline{w}\\ \overline{a}\\ R_m^{-1}\overline{b}\\ R_m^{-1}\overline{c}\\  R^{-1}R_m^{-1}\overline{q}  \end{array} \right]
+\epsilon R^{-1/2}R_m^{-1/2} \left \{ e^{i \alpha x}
\left[ \begin{array}{c} \widetilde{u}\\ \widetilde{v}\\ \widetilde{w}\\ \widetilde{a}\\ \widetilde{b}\\ \widetilde{c}\\\widetilde{q}\end{array} \right]
+\text{c.c.}\label{expincomp}
\right \}.
\end{eqnarray}
Here, the coefficients $\overline{\mathbf{v}}, \overline{\mathbf{b}},\overline{q}$ are real functions of $y,z$, and we call them the vortex components.
The coefficients $\widetilde{\mathbf{v}}, \widetilde{\mathbf{b}},\widetilde{q}$ depend on $y,z$ as well but they are complex and are called the wave components. The c.c. stands for complex conjugate.
Note that in (\ref{expincomp}), we only display the terms relevant to derive the leading order equations. The full asymptotic expansion must, of course, contain the higher-order vortex or wave terms with higher streamwise harmonics, but after we derive the leading order system, we can check that the presence of those higher order terms does not modify the leading order system.
The parameter $\epsilon$ is the wave amplitude to be fixed in terms of $R$ later and is, meanwhile, assumed to be of $O(1)$ or smaller.

Substituting (\ref{expincomp}) to (\ref{incompgov}) and then neglecting some small terms, we have the vortex equations
\begin{subequations}\label{invvovo}
\begin{eqnarray}
~[P_m^{-1}(\overline{v}\partial_y +\overline{w}\partial_z)-\triangle_2]
\left[ \begin{array}{c} \overline{u}\\ \overline{v}\\ \overline{w} \end{array} \right]
+\left[ \begin{array}{c} 0\\ \overline{q}_y\\ \overline{q}_z \end{array} \right]
=
 \left[ \begin{array}{c} -U_b'\overline{v} +(\overline{b}\partial_y +\overline{c}\partial_z)\overline{a}\\ P_m^{-1}(\overline{b}\partial_y +\overline{c}\partial_z)\overline{b}\\ P_m^{-1}(\overline{b}\partial_y +\overline{c}\partial_z)\overline{c} \end{array} \right]\nonumber \\
 -\epsilon^2\left[ \begin{array}{c} 0\\ 
\{ (|\widetilde{v}|^2-|\widetilde{b}|^2 )_y +( \widetilde{v} \widetilde{w}^* -\widetilde{b} \widetilde{c}^*)_z \}+\text{c.c.} \\ 
\{  (|\widetilde{w}|^2-|\widetilde{c}|^2 )_z + ( \widetilde{v} \widetilde{w}^* -\widetilde{b} \widetilde{c}^*)_y \}+\text{c.c.}
\end{array} \right],~~~~~\label{invmo}
\\
~[(\overline{v}\partial_y +\overline{w}\partial_z)-\triangle_2]
\left[ \begin{array}{c} \overline{a}\\ \overline{b}\\ \overline{c} \end{array} \right]
-[\overline{b}\partial_y +\overline{c}\partial_z]
\left[ \begin{array}{c} 0\\ \overline{v}\\ \overline{w} \end{array} \right]=
\left[ \begin{array}{c} U_b' \overline{b}+P_m^{-1} (\overline{b}\partial_y +\overline{c}\partial_z)\overline{u}\\ \epsilon^2 \textcolor{black}{P_m}(\widetilde{c}\widetilde{v}^*-\widetilde{b}\widetilde{w}^*)_z+\text{c.c.} \\ -\epsilon^2 \textcolor{black}{P_m} (\widetilde{c}\widetilde{v}^*-\widetilde{b}\widetilde{w}^*)_y+\text{c.c.} \end{array} \right],~~~~~\label{invind}
\\
\overline{v}_y+\overline{w}_z=0, \qquad \overline{b}_y+\overline{c}_z=0, ~~~~~~~\label{invcontv}
\end{eqnarray}
\textcolor{black}{from the $x$-averaged part}, and the wave equations 
\begin{eqnarray}
\left \{
Ui\alpha 
\left[ \begin{array}{c} \widetilde{u}\\ \widetilde{v}\\ \widetilde{w} \end{array} \right] 
+
\left[ \begin{array}{c} \widetilde{v}U_y+\widetilde{w}U_z\\ 0\\ 0 \end{array} \right]
\right \} \hspace{60mm}\nonumber \\
-
\left \{
\overline{a}i\alpha
\left[ \begin{array}{c} \widetilde{a}\\ \widetilde{b}\\ \widetilde{c} \end{array} \right] 
+
\left[ \begin{array}{c} \widetilde{b}\overline{a}_y+\widetilde{c}\overline{a}_z\\ 0\\ 0 \end{array} \right]
\right \}
+\left[ \begin{array}{c} i\alpha \widetilde{q} \\ \widetilde{q}_y \\ \widetilde{q}_z \end{array} \right]=R^{-1}\triangle 
\left[ \begin{array}{c} \widetilde{u}\\ \widetilde{v}\\ \widetilde{w} \end{array} \right],~~~\label{Cwavemo}
\\
\left \{
Ui\alpha
\left[ \begin{array}{c} \widetilde{a}\\ \widetilde{b}\\ \widetilde{c} \end{array} \right] 
+
\left[ \begin{array}{c} \widetilde{v}\overline{a}_y+\widetilde{w}\overline{a}_z\\ 0\\ 0 \end{array} \right]
\right \} \hspace{70mm} \nonumber \\
-
\left \{
\overline{a}i\alpha
\left[ \begin{array}{c} \widetilde{u}\\ \widetilde{v}\\ \widetilde{w} \end{array} \right] 
+
\left[ \begin{array}{c} \widetilde{b}U_y+\widetilde{c}U_z\\ 0\\ 0 \end{array} \right]
\right \}
=R_m^{-1}\triangle 
\left[ \begin{array}{c} \widetilde{a}\\ \widetilde{b}\\ \widetilde{c} \end{array} \right],~~~\label{Cwaveind}\\
i\alpha \widetilde{u}+\widetilde{v}_y+\widetilde{w}_z=0,\qquad i\alpha \widetilde{a}+\widetilde{b}_y+\widetilde{c}_z=0,~~~\label{Cwavecont}
\end{eqnarray}
\end{subequations}
\textcolor{black}{from the fluctuating part in $x$. Here}
 $\triangle_2=\partial_y^2+\partial_z^2$, $\triangle=\triangle_2-\alpha^2$ and 
\begin{eqnarray}
U=U_b+P_m^{-1}\overline{u}-s
\end{eqnarray}
is the doppler-shifted hydrodynamic streak.
In the above equations, we \textcolor{black}{have retained} the diffusion terms in (\ref{Cwavemo}), (\ref{Cwaveind}) because those terms become important within the thin `dissipative layers' surrounding the singularity, \textcolor{black}{about which} we shall explain shortly. Outside the layers, we can formally neglect those diffusion terms.

In order to elucidate the motivation of the scaling (\ref{expincomp}), it is convenient to adopt the terminology introduced by Waleffe (1997) \textcolor{black}{for both of the hydrodynamic and magnetic components;} we call the $x$- and $(y,z)$-components of the vortex part as the streak and roll, respectively. \textcolor{black}{Note that the roll-streak (or vortex) and the wave parts correspond to the mean field and fluctuations in dynamo theory, respectively.}
The $O(R_m^{-1})$ roll scale is chosen so that the viscous-convective balance in the roll-streak equations is achieved.
On the other hand, the wave part is predominantly convected by the streak component (including the base flow), and, therefore, this part can be approximated by ideal flows almost everywhere. 
The wave equations can be regarded as the linear stability problem of the streak, and thus, only the monochromatic wave neutral to the streak can be excited to leading order (this is the reason why there \textcolor{black}{is} only one Fourier mode in (\ref{expincomp})).
From the form of the roll equations, we can find the mechanism by which the wave field causes the feedback to the roll field.
In the momentum equations, the feedback terms are produced by the advection terms and the Lorentz force terms, hereafter called the wave Reynolds stress and the Maxwell stress, respectively (see (\ref{invmo})). 
The similar feedback terms in the induction equations are called the wave electromotive force terms (see (\ref{invind})).

If the diffusion terms are neglected, the wave equations (\ref{Cwavemo})--(\ref{Cwavecont})
are combined to yield the single pressure equation
\begin{eqnarray}
\left ( \frac{\widetilde{q}_y}{\Lambda} \right )_y
+\left ( \frac{\widetilde{q}_z}{\Lambda} \right )_z
-\alpha^2\frac{\widetilde{q}}{\Lambda}
=0,\qquad \Lambda=U^2-\overline{a}^2.\label{incompinvwave}
\end{eqnarray}
The pressure equation is the natural generalisation \textcolor{black}{for the one} derived in the previous resonant absorption studies \textcolor{black}{Sakurai et al. (1991), Goossens et al. (1995)} for one-dimensional background flows. \textcolor{black}{Those works are concerned with compressible flows; so, incompressible limit must be compared with the above equation} - see the review by \textcolor{black}{Goossens et al.} (2011) also.
Given the hydrodynamic and magnetic streaks, the equation (\ref{incompinvwave}) constitutes the linear eigenvalue problem for the eigenvalue $s$. 
\textcolor{black}{The pressure eigensolution may become singular when $\Lambda$ vanishes, namely when the doppler-shifted hydrodynamic and magnetic streak energies are equipartitioned (i.e. $U^2= \overline{a}^2$).
Hereinafter, we denote the singular positions as $y=f_{\pm}(z)$ and assume that $U\pm \overline{a}=0$ is satisfied there. 
Those two types of singularities correspond to the Alfv\'en resonant points discussed in the resonant absorption theory. }


Once $\widetilde{q}$ is solved, we can express the other wave components as
\begin{subequations}\label{incompinvwave2}
\begin{eqnarray}
\widetilde{v}=-\frac{\widetilde{q}_yU}{i\alpha \Lambda},\qquad \widetilde{w}=-\frac{\widetilde{q}_zU}{i\alpha \Lambda},\qquad
\widetilde{b}=-\frac{\widetilde{q}_y\overline{a}}{i\alpha \Lambda},\qquad \widetilde{c}=-\frac{\widetilde{q}_z\overline{a}}{i\alpha \Lambda},\label{invvwbc}\\
\widetilde{u}=-\frac{1}{\Lambda}\left (\frac{\widetilde{q}_yU_y+\widetilde{q}_zU_z}{\alpha^2}+U\widetilde{q} \right ),\qquad 
\widetilde{a}=-\frac{1}{\Lambda}\left (\frac{\widetilde{q}_y\overline{a}_y+\widetilde{q}_z\overline{a}_z}{\alpha^2}+\overline{a}\widetilde{q} \right ).
\end{eqnarray}
\end{subequations}
Here, we remark that the behaviour of the cross-stream components of the magnetic wave is simply related to the hydrodynamic counterpart as
\begin{eqnarray}
(\widetilde{b},\widetilde{c})=\frac{\overline{a}}{U}(\widetilde{v},\widetilde{w}).\label{frozen}
\end{eqnarray}
This is due to the absence of the diffusion in the wave induction equations and analogues to the Alfv\'en's frozen-in theorem. 
\textcolor{black}{Hereafter,} we say that the magnetic wave is `frozen' to the hydrodynamic wave when (\ref{frozen}) holds.

On making use of \textcolor{black}{the inviscid wave solutions (\ref{incompinvwave2})} to (\ref{invvovo}), we have the outer roll-streak equations
\begin{subequations}\label{outvortex}
\begin{eqnarray}
~[P_m^{-1}(\overline{v}\partial_y +\overline{w}\partial_z)-\triangle_2]
\left[ \begin{array}{c} \overline{u}\\ \overline{v}\\ \overline{w} \end{array} \right]
+\left[ \begin{array}{c} 0\\ \overline{q}_y\\ \overline{q}_z \end{array} \right]
=
 \left[ \begin{array}{c} -U_b'\overline{v} +(\overline{b}\partial_y +\overline{c}\partial_z)\overline{a}\\ P_m^{-1}(\overline{b}\partial_y +\overline{c}\partial_z)\overline{b}\\ P_m^{-1}(\overline{b}\partial_y +\overline{c}\partial_z)\overline{c} \end{array} \right]\nonumber \\
 -\frac{\epsilon^2}{\alpha^2}\left[ \begin{array}{c} 0\\ 
\{ (\Lambda^{-1}|\widetilde{q}_y|^2 )_y +( \Lambda^{-1}\widetilde{q}_y\widetilde{q}_z^* )_z \}+\text{c.c.} \\ 
\{  (\Lambda^{-1}|\widetilde{q}_z|^2  )_z + ( \Lambda^{-1}\widetilde{q}_y\widetilde{q}_z^*)_y \}+\text{c.c.}
\end{array} \right],~~~\label{epsfeedback}
\\
~[(\overline{v}\partial_y +\overline{w}\partial_z)-\triangle_2]
\left[ \begin{array}{c} \overline{a}\\ \overline{b}\\ \overline{c} \end{array} \right]
-[\overline{b}\partial_y +\overline{c}\partial_z]
\left[ \begin{array}{c} 0\\ \overline{v}\\ \overline{w} \end{array} \right]=
\left[ \begin{array}{c} U_b' \overline{b}+P_m^{-1} (\overline{b}\partial_y +\overline{c}\partial_z)\overline{u}\\ 0 \\ 0
\end{array} \right],~~~\label{zerofeedback}\\
\overline{v}_y+\overline{w}_z=0, \qquad \overline{b}_y+\overline{c}_z=0. ~~~~~~~~~~~~~~~~~~
\end{eqnarray}
\end{subequations} 
There is actually no wave electromotive force terms in the above outer induction equations. 
\textcolor{black}{
The reason is that, as remarked earlier, the cross-streamwise components of hydrodynamic and magnetic inviscid wave solutions must behave similar way except for the factor proportional to $U$ or $\overline{a}$; see (\ref{invvwbc}).
That frozen wave ensures the cancellation of the two nonlinear terms in the roll parts of the induction equations.
In other words, if there is a feedback effect from the wave to the magnetic roll, the wave must be resistive.}

\textcolor{black}{
Thus in the next sections we analyse the dissipative layer surrounding the singular curve of the wave problem to 
find some possible discontinuities in the roll field. That `jump' represents the feedback mechanism from the wave to the roll-streak flow, and together with this condition, the wave equation (\ref{incompinvwave}) and the vortex equations (\ref{outvortex}) form a closure. 
For plane Couette flow, the wave equation (\ref{incompinvwave}) can be solved together with the boundary conditions $\widetilde{q}_y=0$ at $y=\pm 1$, which ensures that the walls are impermeable 
(Near wall boundary layers must be inserted to satisfy the other boundary conditions but the dynamics in the layer is passive so the analysis of it is not necessary). The roll-streak part is fully viscous so the no-slip conditions on the wall must be applied. 
}

\textcolor{black}{
As mentioned earlier the present framework can also be used to describe the large Reynolds number limit of some shearing box solutions by modifying the boundary conditions. 
The conditions for the roll-streak flow can be found straightforwardly because, excluding the base part, they should be periodic in $y$. 
At first glance the condition for the wave part is not simple because there are many waves of different speeds in the problem. 
However Hall (2018) showed that the stability problem of each wave can be decoupled at the asymptotic limit, and the amplitude of it should decay rapidly away from the critical point;
hence $|\widetilde{q}|\rightarrow 0$ as $|y|\rightarrow \infty$ is the appropriate boundary conditions for the wave component.
}

\vspace{5mm}
\subsection{Body-fitted coordinate}

\begin{figure}
\centering
\includegraphics[scale=0.5,clip]{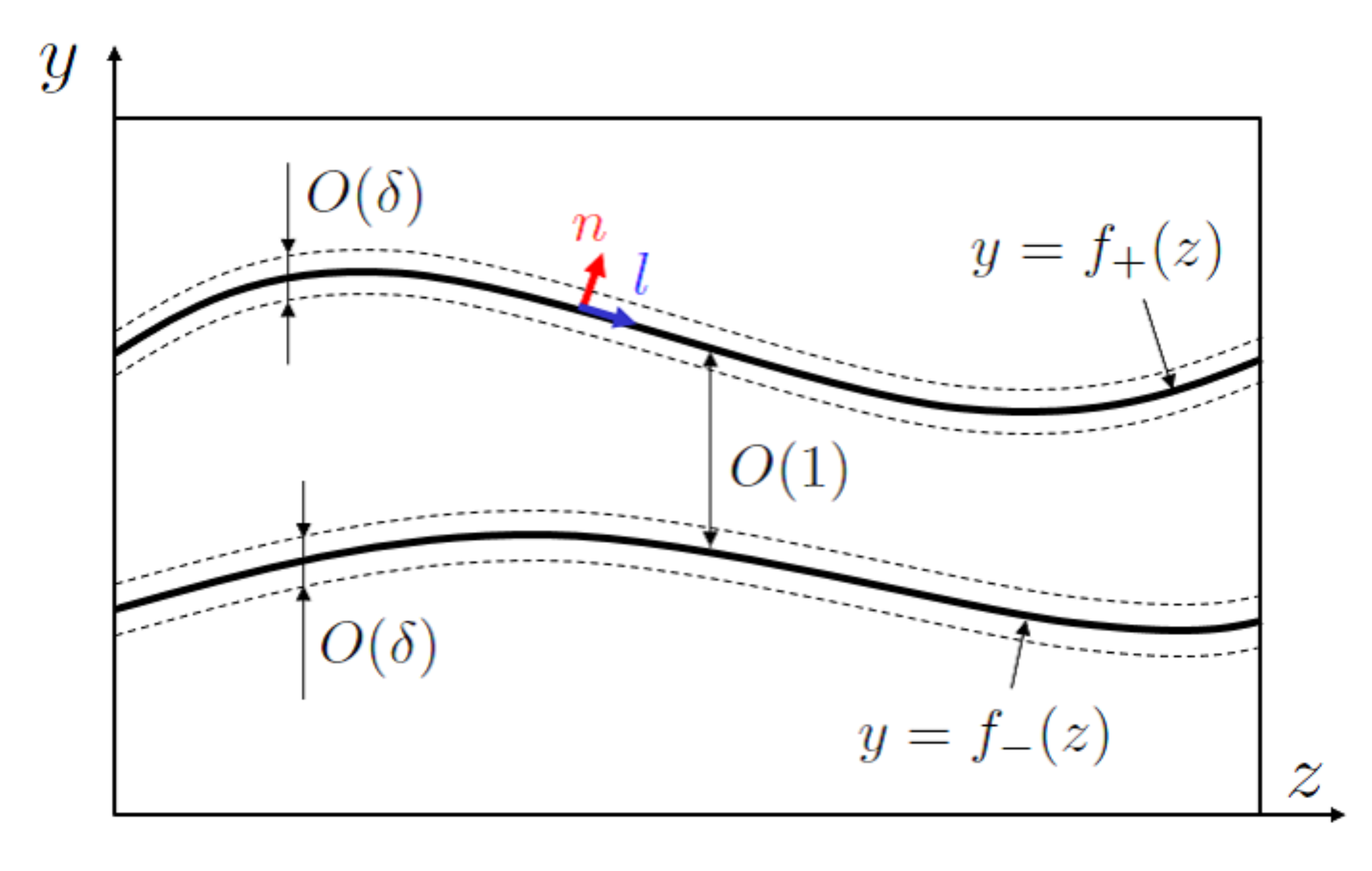}
\label{fig:bif}
\caption{
Schematic of type 1 interaction studied in section 3.2. The body-fitted coordinate $(n,l)$ is shown for the upper resonant curve $y=f_+(z)$. \textcolor{black}{The dissipative layer thickness $\delta$ is defined in (\ref{thickness}).}
Type 2 interaction occurs when the distance between the two resonant curves becomes $O(\delta)$; see section 3.3.
}
\end{figure}

One of the major aims \textcolor{black}{of} the resonant absorption theory is to find the connection formulae, that tells us how to connect the outer solutions across the resonant point. 
Here, similar formulae must be found from the inner analysis to find the jump \textcolor{black}{that} completes the nonlinear system at large $R$ with the outer problem. 
However, the singularity now occurs on a curve in $y$--$z$ plane rather than a point so it is convenient to use `body-fitted' \textcolor{black}{coordinates} attached to the resonant curve \textcolor{black}{(Slattery 1999).
Let us consider a curve $y=f(z)$ and write the length measured along straight lines that are normal to this curve as $n$, and the arc length measured along the curve as $l$; see figure 2.
Any location in $y$--$z$ plane can be specified by the position vector 
\begin{eqnarray}
\mathbf{r}=\mathbf{r}_0+n\mathbf{e}_n,
\end{eqnarray}
where $\mathbf{r_0}=f\mathbf{e}_y+z\mathbf{e}_z$ is a point on the curve $y=f(z)$,specified by the arc length $l$. (Here $\mathbf{e}_n, \mathbf{e}_y, \mathbf{e}_z$, and $\mathbf{e}_x, \mathbf{e}_l$ to be used later are all unit vectors.)
We further note that by definition of the $n$--$l$ coordinates, 
\begin{eqnarray}
\mathbf{e}_n=\frac{dz}{dl}(\mathbf{e}_y-f'\mathbf{e}_z),\qquad \frac{dz}{dl}=\frac{1}{\sqrt{1+f'^2}}.
\end{eqnarray}
Hence a straightforward calculation yields the Lam\'e coefficients $|\frac{\partial \mathbf{r}}{\partial n}|=1$ and $|\frac{\partial \mathbf{r}}{\partial l}|=g$, where
\begin{eqnarray}
g(n,l)=1+\chi n,\qquad \chi(l)=-\frac{f''}{(1+f'^2)^{3/2}}.
\end{eqnarray}
Note that $\chi$ is the curvature of the critical curve $y=f(z)$.
Given those Lam\'e coefficients, we can use the standard orthogonal curvilinear coordinate theory to derive the body-fitted coordinate expression of the equations.
Writing the velocity and magnetic field vectors as
\begin{eqnarray}
\mathbf{v}=u(x,n,l)\mathbf{e}_x+\mathcal{V}(x,n,l)\mathbf{e}_n+\mathcal{W}(x,n,l)\mathbf{e}_l,\\
\mathbf{b}=a(x,n,l)\mathbf{e}_x+\mathcal{B}(x,n,l)\mathbf{e}_n+\mathcal{C}(x,n,l)\mathbf{e}_l,
\end{eqnarray}
we can transform the governing equations (\ref{incompgov}) into}
\begin{subequations}\label{govbody}
\begin{eqnarray}
~[(u-s)\partial_x+\mathcal{V}\partial_n+\frac{\mathcal{W}}{g}\partial_l]
\left[ \begin{array}{c} u\\ \mathcal{V}\\ \mathcal{W} \end{array} \right]
-[a\partial_x+\mathcal{B}\partial_n+\frac{\mathcal{C}}{g}\partial_l]
\left[ \begin{array}{c} a\\ \mathcal{B}\\ \mathcal{C} \end{array} \right]
+\frac{g_n}{g}
\left[ \begin{array}{c} 0\\ (\mathcal{C}^2-\mathcal{W}^2)\\ (\mathcal{V}\mathcal{W}-\mathcal{B}\mathcal{C}) \end{array} \right]\nonumber \\
=-\left[ \begin{array}{c} q_x\\ q_n\\ g^{-1}q_l \end{array} \right]
+R^{-1}  \left[ \begin{array}{c} L_1u\\ L_2\mathcal{V}-L_3\mathcal{W}\\ L_2\mathcal{W}+L_3\mathcal{V} \end{array} \right], ~~~~~~~
\\
~[(u-s)\partial_x+\mathcal{V}\partial_n+\frac{\mathcal{W}}{g}\partial_l]
\left[ \begin{array}{c} a\\ \mathcal{B}\\ \mathcal{C} \end{array} \right]
-[a\partial_x+\mathcal{B}\partial_n+\frac{\mathcal{C}}{g}\partial_l]
\left[ \begin{array}{c} u\\ \mathcal{V}\\ \mathcal{W}  \end{array} \right]
+\frac{g_n}{g}
\left[ \begin{array}{c} 0\\ 0\\ (\mathcal{B}\mathcal{W}-\mathcal{V}\mathcal{C}) \end{array} \right]\nonumber \\=
R_m^{-1} \left[ \begin{array}{c} L_1a\\ L_2\mathcal{B}-L_3\mathcal{C}\\ L_2\mathcal{C}+L_3\mathcal{B} \end{array} \right], ~~~~~~~
 \\
u_x+g^{-1}(g\mathcal{V})_n+g^{-1}\mathcal{W}_l=0,\qquad a_x+g^{-1}(g\mathcal{B})_n+g^{-1}\mathcal{C}_l=0,~~~~~~~~
\end{eqnarray}
where
\begin{eqnarray}
L_1(~)=(~)_{xx}+g^{-1}(g(~)_n)_n+g^{-1}(g^{-1}(~)_l)_l,\\
L_2(~)=(~)_{xx}+(g^{-1}(g(~))_n)_n+g^{-1}(g^{-1}(~)_l)_l,\\
L_3(~)=g^{-1}(g^{-1}(g(~))_n)_l-(g^{-1}(~)_l)_n.
\end{eqnarray}
\end{subequations}

Then for $\mathcal{V}, \mathcal{W}, \mathcal{B}, \mathcal{C}$ we can also apply the vortex-wave decomposition 
\begin{eqnarray}
\left[ \begin{array}{c}  \mathcal{V}\\ \mathcal{W}\\\mathcal{B}\\\mathcal{C} \end{array} \right]=
\left[ \begin{array}{c}  R_m^{-1}\overline{\mathcal{V}}\\ R_m^{-1}\overline{\mathcal{W}}\\ R_m^{-1}\overline{\mathcal{B}}\\ R_m^{-1}\overline{\mathcal{C}}  \end{array} \right]
+\epsilon R^{-1/2}R_m^{-1/2} \left \{ e^{i \alpha x}
\left[ \begin{array}{c} \widetilde{\mathcal{V}}\\ \widetilde{\mathcal{W}}\\ \widetilde{\mathcal{B}}\\ \widetilde{\mathcal{C}}\end{array} \right]
+\text{c.c.}
\right \}\label{vwbodyfit}
\end{eqnarray}
to find the body-fitted coordinate version of the vortex and wave equations.
The local behaviour of the outer solution near the resonant curve gives the matching conditions needed to solve the solutions within the dissipative layer. 
Near the resonant curve, the wave behaves \textcolor{black}{in} a singular way, so its amplitude must be increased.
It is this amplified inner wave that creates derivative jumps in the outer roll components.
The aim of the next two subsections is to derive the analytic form of the jump, by the generalisation of the connection formulae obtained in the previous resonant point analysis.
The wave Reynolds-Maxwell stress and the roll stress within the dissipative layer must be in balance, and \textcolor{black}{this} condition gives the appropriate size of the wave amplitude $\epsilon$ in terms of our intrinsic small parameter $1/R$.
We shall show that the magnitude of $\epsilon$ changes depending on the distance between the two resonant layers.

\vspace{5mm}

\subsection{Type 1: the connection formulae for the case $(f_+-f_-)\sim O(1)$}

\begin{figure}
\centering
\includegraphics[scale=0.4,clip]{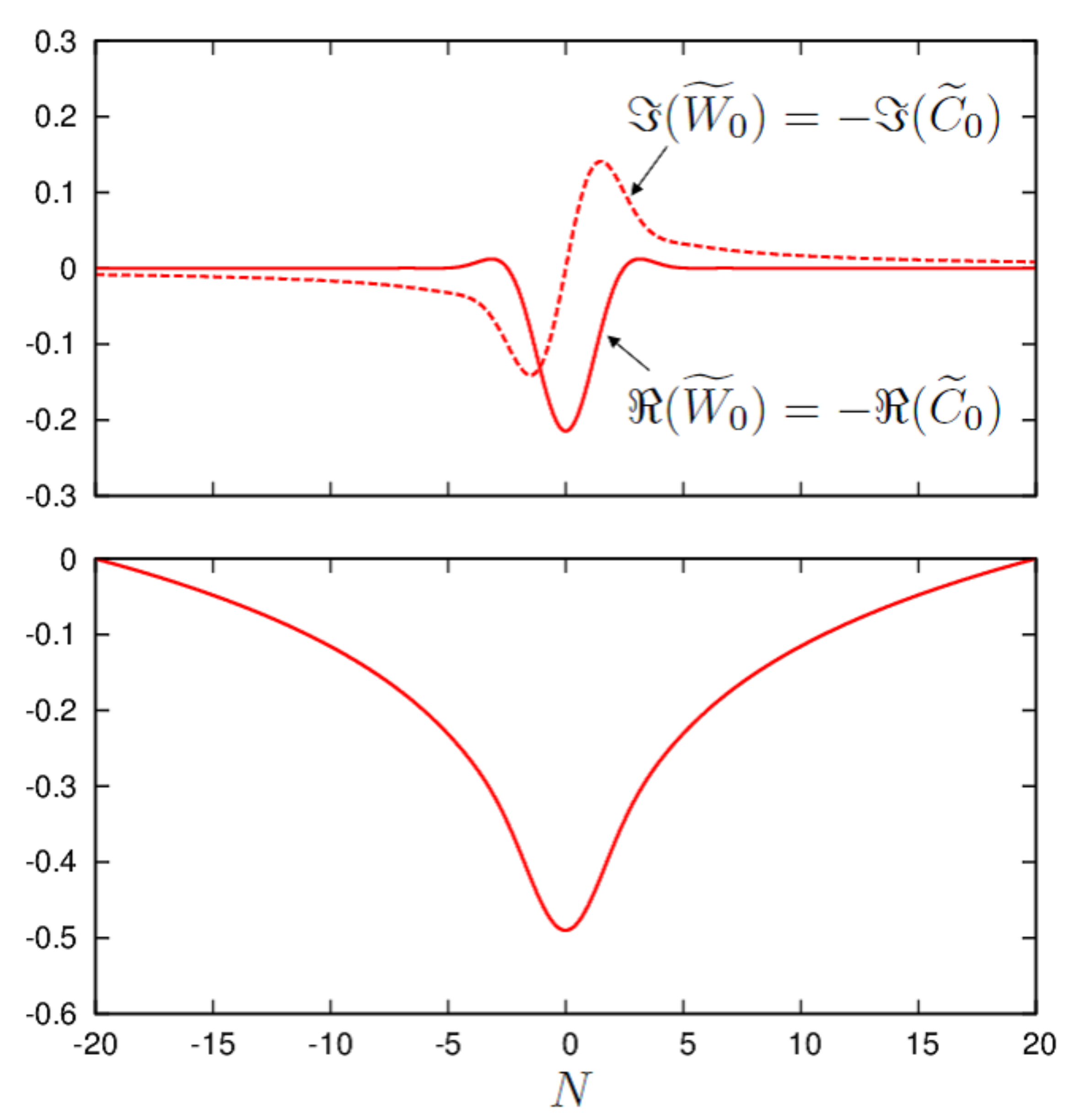}
\label{fig:bif}
\caption{
The upper panel shows the inner wave solution for type 1 interaction (\ref{incompwsol}). The representative parameters $(\alpha,\lambda_0,\lambda_u,\lambda_a,\widetilde{Q}_0',P_m)=(1,1,2,1,1,1/5)$ are chosen for $y=f_+(z)$.
The lower panel is the \textcolor{black}{force} from the inner wave to the hydrodynamic roll component (\ref{inttype1}).
}
\end{figure}

In this section, we assume the distance between the two resonant curves is $O(1)$; this case is called type 1 interaction and \textcolor{black}{the} schematic of it is shown in figure 2.
Using the body-fitted coordinate near the resonant curve $y=f_{\pm}(z)$, we can Taylor expand the streak as
\begin{eqnarray}
U=\lambda_0(l)+\lambda_u(l) n+\cdots,\qquad
\overline{a}=\mp \lambda_0(l)+\lambda_a(l) n+\cdots.\label{expUA}
\end{eqnarray}
This means that we can write
\begin{eqnarray}
\Lambda=\lambda(l) n+\cdots,\qquad
\lambda=2\lambda_0 (\lambda_u\pm \lambda_a ),
\end{eqnarray}
where we assume $\lambda(l) \neq 0$ for all $l$. 
Of course $n=0$ is the resonant curve, and here, $\Lambda$ vanishes as required. 
The singularity there must be resolved within the thin dissipative layers surrounding $y=f_{\pm}(z)$, where we retain some diffusion effects.
The flows inside and outside of the dissipative layer analysed separately are then connected through the matching conditions.

Now let us analyse the singular behaviour of the outer wave solution at the edge of the dissipative layer. 
In the body-fitted coordinates, the inviscid wave equation (\ref{incompinvwave}) becomes
\begin{equation}
\frac{\partial^{2}\widetilde{q}}{\partial n^{2}} +\frac{\chi}{g}\frac{\partial
\widetilde{q}}{\partial n} +\frac{1}{g^{2}}\frac{\partial^{2}\widetilde{q}}{\partial l^{2}}
-\frac{\chi^{\prime}n}{g^{3}}\frac{\partial \widetilde{q}}{\partial l}
-\alpha^{2}\widetilde{q} -\frac{1}{\Lambda} \left( \frac{\partial \Lambda }{\partial n}\frac{\partial
\widetilde{q}}{\partial n}+\frac{1}{g^{2}}\frac{\partial \Lambda}{\partial l}%
\frac{\partial \widetilde{q}}{\partial l} \right)  =0.\label{invwavebody}
\end{equation}
Since $\Lambda \sim O(n)$ when $n$ is small, the above equation suggests the Frobenius expansion 
\begin{eqnarray}
\widetilde{q}=\widetilde{q}_0(l)+\widetilde{q}_{2L}(l)n^2 (\ln |n|+i\Theta)+\widetilde{q}_2(l)n^2+\cdots,\label{incomplog}
\end{eqnarray}
where
\begin{eqnarray}
\Theta=
\left \{
\begin{array}{c}
0 \qquad \text{if}\qquad n>0, \\
\theta  \qquad \text{if}\qquad n<0.
\end{array}
\right .\label{logthe}
\end{eqnarray}
The quantity $\theta$ appeared in (\ref{logthe}) is the jump associated with the logarithmic singularity and physically represents the phase shift of the wave across the resonant curve. 
The jump is induced by the diffusion, so the value of $\theta$ should be estimated by the solution within the dissipative layer \textcolor{black}{(it will be found later that the wave in the layer is linear and the modulus of $\theta$ is the well-known value of $\pi$).}
We can fix the coefficients $\widetilde{q}_0, \widetilde{q}_2$ from the boundary conditions, and all the other coefficients in the expansion can be fixed in terms of those two coefficients. 
In particular, \textcolor{black}{on} substituting (\ref{incomplog}) to (\ref{invwavebody}), we can find
\begin{eqnarray}
\widetilde{q}_{2L}=\frac{1}{2}\left (\alpha^2 \widetilde{q}_0+\frac{\lambda'}{\lambda} \widetilde{q}_{0}'-\widetilde{q}_0'' \right ).
\end{eqnarray}

From the body-fitted coordinate expression of (\ref{incompinvwave2}) and (\ref{incomplog}), we can find the small $n$ asymptotic behaviours of the other wave components
\begin{eqnarray}
\widetilde{u},\widetilde{a} \sim O(n^{-1}),\qquad
\widetilde{\mathcal{V}},\widetilde{\mathcal{B}} \sim O(\ln |n|), \qquad 
\widetilde{\mathcal{W}},\widetilde{\mathcal{C}} \sim O(n^{-1}).\label{sininc}
\end{eqnarray}
More precisely, the cross-stream components expand \textcolor{black}{as}
\begin{subequations}\label{limitwave}
\begin{eqnarray}
\widetilde{\mathcal{V}}=-\frac{\lambda_0 2\widetilde{q}_{2L}\ln |n|+2(\widetilde{q}_2+i\Theta \widetilde{q}_{2L})}{i\alpha \lambda}+\cdots,\\
\widetilde{\mathcal{B}}=\pm \frac{\lambda_0 2\widetilde{q}_{2L}\ln |n|+2(\widetilde{q}_2+i\Theta \widetilde{q}_{2L})}{i\alpha \lambda}+\cdots,\\
\widetilde{\mathcal{W}}=-\frac{\lambda_0 \widetilde{q}_0'}{i\alpha \lambda n}+\cdots,\qquad \widetilde{\mathcal{C}}=\pm \frac{\lambda_0 \widetilde{q}_0'}{i\alpha \lambda n}+\cdots.
\end{eqnarray}
\end{subequations}

Using the stretched normal coordinate $N=n/\delta$ with the dissipative layer thickness 
\textcolor{black}{
\begin{eqnarray}
\delta=R^{-1/3}, \label{thickness}
\end{eqnarray}
}
the inner expansions consistent to (\ref{incomplog}), (\ref{sininc}) can be found as
\begin{subequations}\label{innerwaveexp}
\begin{eqnarray}
\widetilde{q}=\widetilde{Q}_0(N,l)+(\delta^2 \ln \delta)\widetilde{Q}_{2L}(N,l)+\delta^2\widetilde{Q}_{2}(N,l)+\cdots,~~~\\
\widetilde{u}=\delta^{-1}\widetilde{U}_0(N,l)+\cdots,\qquad \widetilde{a}=\delta^{-1}\widetilde{A}_0(N,l)+\cdots,~~~\\
\widetilde{\mathcal{V}}=(\ln \delta) \widetilde{V}_{0L}(N,l)+\widetilde{V}_0(N,l)+\cdots,~~~ \widetilde{\mathcal{B}}=(\ln \delta) \widetilde{B}_{0L}(N,l)+\widetilde{B}_0(N,l)+\cdots,~~~\\
\widetilde{\mathcal{W}}=\delta^{-1}\widetilde{W}_0(N,l)+\widetilde{W}_1(N,l)+\cdots,~~~
\widetilde{\mathcal{C}}=\delta^{-1}\widetilde{C}_0(N,l)+\widetilde{C}_1(N,l)+\cdots.\label{innerwcexp}~~~
\end{eqnarray}
\end{subequations}
The dissipative layer thickness is chosen in such a way that the diffusion terms influence the inner flow.

The inner wave equations are obtained by substituting (\ref{expUA}), (\ref{innerwaveexp}) to the body-fitted coordinate version of the wave equations. We must analyse the equations up to the first order.
From the $n$-components we can find that 
\begin{subequations}\label{subwww}
\begin{eqnarray}
\widetilde{Q}_{0N}=0,\label{eqvqq}\\
\widetilde{V}_0\pm \widetilde{B}_0=0,\label{eqv00}
\end{eqnarray}
whilst the $l$-components yield
\begin{eqnarray}
\widetilde{W}_0\pm \widetilde{C}_0=0,\label{infrozen}\\
i\alpha \lambda_0 (\widetilde{W}_1\pm \widetilde{C}_1)+i\alpha N(\lambda_u \widetilde{W}_0-\lambda_a \widetilde{C}_0)=-\widetilde{Q}_{0l}+\widetilde{W}_{0NN},\label{subwww1}\\
\pm i\alpha \lambda_0 (\widetilde{W}_1\pm \widetilde{C}_1)+i\alpha N(\lambda_u \widetilde{C}_0-\lambda_a \widetilde{W}_0)=P_m^{-1}\widetilde{C}_{0NN}.\label{subwww2}
\end{eqnarray}
\end{subequations}
From (\ref{eqv00}) and (\ref{infrozen}), we see that the zeroth order components remain frozen within the dissipative layer.

\textcolor{black}{Equations} (\ref{infrozen})-(\ref{subwww2}) can \textcolor{black}{then} be combined to yield the single equation for $\widetilde{W}_0$
\begin{eqnarray}
i\alpha \frac{\lambda}{\lambda_0} N \widetilde{W}_0=-\widetilde{Q}_{0}'+(1+P_m^{-1})\widetilde{W}_{0NN}.\label{WWW}
\end{eqnarray}
Here $\widetilde{Q}_{0}$ is a function of $l$ only from (\ref{eqvqq}) and the prime denotes \textcolor{black}{an} ordinary differentiation.
The equation is linear and the solution $\widetilde{W}_0$ can be found analytically. Note that the $l$-component of the wave is the most singular and it is the only component necessary to estimate the jump in the roll components.
 
We note in passing that in \textcolor{black}{our three dimensional theory} we can safely assume that there are no significant effects \textcolor{black}{from} the nonlinear terms with respect to the waves, \textcolor{black}{unlike two-dimensional flows.}
For two-dimensional \textcolor{black}{cases}, the resonant absorption theories have been extended to include the wave nonlinearity within the dissipation layer. 
The nonlinear effects around the Alfv\'en resonant point is analysed by \textcolor{black}{Ruderman \& Goossens (1993), Clack et al. (2009) and Ruderman et al. (2010); see the review by Ballai \& Ruderman (2011) also. 
(The absorption theories usually concerns compressible flows where the singularity also occurs at the cusp wave resonant point. The corresponding nonlinear equations within the dissipation layer can be found in Ruderman et al. (1997), Ballai \& Erdelyi (1998), Clack \& Ballai (2008).)
In those works it was found that we need certain size of the wave amplitude to balance the nonlinear term.}
However, in our case, the appropriate wave amplitude must be fixed in order to drive the roll component, as we shall see shortly. Therefore, we cannot vary it to balance the nonlinear terms in the dissipative layer wave equations as in the previous studies. \textcolor{black}{The outer wave amplitude to be shown is of $O(R^{-1})$ or smaller, by which we can safely conclude that the dissipation layer is linear.}
\textcolor{black}{A} similar conclusion was obtained in the purely hydrodynamic studies. 
Hall \& Smith (1991) found that the inner wave equations in their theory should be always linear \textcolor{black}{for} the same reason, although if the flow is two-dimensional, \textcolor{black}{the} so-called nonlinear critical layer is possible (Smith \& Bodonyi 1982; Deguchi \& Walton 2018). 

Using the new variable 
\begin{eqnarray}
\zeta=\text{sgn}(\lambda \lambda_0)\left (\frac{i\alpha |\lambda|}{|\lambda_0|(1+P_m^{-1})} \right )^{1/3}N,
\end{eqnarray}
the equation (\ref{WWW}) can be transformed into
\begin{eqnarray}
\left (\frac{\partial^2}{\partial \zeta^2}-\zeta \right )\widetilde{W}_0=\left (\frac{|\lambda_0|}{i\alpha |\lambda|} \right )^{2/3}\frac{Q_0'}{(1+P_m^{-1})^{1/3}}.
\end{eqnarray}
The solution of this equation \textcolor{black}{that is matching} the outer solution can be found as
\begin{eqnarray}
\widetilde{W}_0=\left (\frac{|\lambda_0|}{i\alpha |\lambda|} \right )^{2/3}\frac{Q_0'}{(1+P_m^{-1})^{1/3}}S(\zeta),\label{incompwsol}
\end{eqnarray}
where
\begin{eqnarray}
S(\zeta)\equiv-i^{2/3}\int^{\infty}_0 e^{-\frac{t^3}{3}-i^{2/3}\zeta t}dt \label{defS}
\end{eqnarray}
satisfies $S''-\zeta S=1$ and $S\rightarrow -\zeta^{-1}$ in the far-field limit ($|i^{-1/3}\zeta|\rightarrow \infty$).

The integral of $S$ with respect to $N$ 
\begin{eqnarray}
\kappa(\zeta)\equiv\int^{\zeta}_0S(\zeta_*)\zeta_*=\int^{\infty}_0 \frac{e^{-i^{2/3}\zeta t}-1}{t}e^{-\frac{t^3}{3}}dt.\label{defkappa}
\end{eqnarray}
produces a logarithmic function with a constant jump for large $|N|$
\begin{eqnarray}
-\kappa \rightarrow
\left \{
\begin{array}{c}
\ln N_*+c_{00}+\cdots\qquad \text{as}~~~~N_*\rightarrow \infty,\\
\ln |N_*|-i\pi \text{sgn}(\lambda \lambda_0)+c_{00}+\cdots \qquad \text{as}~~~~N_*\rightarrow -\infty,
\end{array}
\right .\label{jumpSco}
\end{eqnarray}
where $c_{00}$ is a constant (see Appendix A.1).
In short, the logarithm and jump are exactly what we saw in the pressure expansion (\ref{incomplog}). 
Thus, in order to match the solution, the value of the logarithmic phase shift in (\ref{logthe}) must be
\begin{eqnarray}
\theta=-\pi \text{sgn}(\lambda \lambda_0).\label{phase1}
\end{eqnarray}
The analogues result was obtained by Sakurai et al. (1991) \textcolor{black}{and Goossens et al. (1995)} in \textcolor{black}{the} solar physics community, although it has long been recognised in the hydrodynamic study; see \textcolor{black}{Lin (1945, 1955),} Haberman (1972) for example.

The other wave components can be explicitly solved, but since they are not necessary to find the jumps in the roll components, we omit further inner wave analysis here. 

\vspace{5mm}

The inner roll fields expand
\begin{subequations}\label{exproll}
\begin{eqnarray}
\overline{q}=\overline{Q}_0(N,l)+\cdots, ~
\overline{\mathcal{V}}=\overline{V}_0(l)+\delta \overline{V}_1(N,l)+\cdots,~
\overline{\mathcal{B}}=\overline{B}_0(l)+\delta \overline{B}_1(N,l)+\cdots,~~~~~~\\
\overline{\mathcal{W}}=\overline{W}_0(l)+\delta \overline{W}_1(N,l)+\cdots,~~~
\overline{\mathcal{C}}=\overline{C}_0(l)+\delta \overline{C}_1(N,l)+\cdots.~~~~~~~~~~
\end{eqnarray}
\end{subequations}
The outer roll velocity and magnetic fields should be continuous across the dissipative layer, but the normal derivative of the tangential components, namely the leading order vorticity and current, might suffer jumps. 

Here, note that there is no jump in the normal derivative of the normal components because from the solenoidal conditions, we have $\overline{W}_0'+\overline{V}_{1N}=0, \overline{C}_0'+\overline{B}_{1N}=0$. 
Note that the condition $[\overline{V}_{1N}]^{\infty}_{-\infty}=[\overline{B}_{1N}]^{\infty}_{-\infty}=0$ (where $[~]^{\infty}_{-\infty}=[~]^{\infty}_{N=-\infty}$) can be written in the outer variable expression $[\overline{\mathcal{V}}_{n}]^{+}_{-}=[\overline{\mathcal{B}}_{n}]^{+}_{-}=0$ (where $[~]^+_-=[~]^{0+}_{n=0-}$).

The vorticity and current jumps can be found from the roll equations. 
Substituting the inner expansions (\ref{innerwaveexp}) and (\ref{exproll}) to the body-fitted coordinate forms of the roll equations and \textcolor{black}{integrating} them over $N\in (\infty, -\infty)$, 
\begin{subequations}\label{jumpp}
\begin{eqnarray}
\epsilon^2 \left \{[|\widetilde{\mathcal{V}}|^2-|\widetilde{\mathcal{B}}|^2]^{+}_{-}+
\chi \int^{\infty}_{-\infty} 2(\widetilde{W}_0^*\widetilde{W}_1-\widetilde{C}_0^*\widetilde{C}_1)dN\right \}+\text{c.c.}+\cdots=[\overline{Q}_{0}]^{\infty}_{-\infty},~~~\\
\epsilon^2  \left \{[\widetilde{\mathcal{V}}\widetilde{\mathcal{W}}^*-\widetilde{\mathcal{B}}\widetilde{\mathcal{C}}^*]^{+}_{-}+\frac{\partial}{\partial l}\int^{\infty}_{-\infty} 2(\widetilde{W}_0^*\widetilde{W}_1-\widetilde{C}_0^*\widetilde{C}_1) dN \right \}+\text{c.c.}+\cdots=[\overline{W}_{1N}]^{\infty}_{-\infty},~~~\\
~\epsilon^2 \textcolor{black}{P_m}[\widetilde{\mathcal{V}}\widetilde{\mathcal{C}}^*-\widetilde{\mathcal{B}}\widetilde{\mathcal{W}}^*]^{+}_{-}+\text{c.c.}+\cdots=[\overline{C}_{1N}]^{\infty}_{-\infty},~~~\label{cccccc}
\end{eqnarray}
\end{subequations}
where 
the superscript asterisks are used to express the complex conjugate.
The first two equations are derived from the $n$- and $l$-components of the hydrodynamic roll equations, and the third equation comes from the $l$-component of the magnetic roll equations (analysis of the other equations are not necessary to find the jumps).
In order to derive the above equations, it is important to note that terms \textcolor{black}{such as} $(|\widetilde{W}_0|^2-|\widetilde{C}_0|^2)$ should vanish in view of the frozen condition (\ref{infrozen}). In order to balance both sides of the equations, we must choose
\begin{eqnarray}
\epsilon=1,
\end{eqnarray}
which is larger than \textcolor{black}{$\epsilon=R^{-1/6}$ in} Hall \& Smith (1991).

We can further show that the integral of $(\widetilde{W}_0^*\widetilde{W}_1-\widetilde{C}_0^*\widetilde{C}_1)+\text{c.c.}$ appearing in (\ref{jumpp}) should also vanish.
The behaviour of the inner wave forcing can be found in figure 3 where
\begin{eqnarray}
J_{N_0}(N,l)=\int^{N}_{-N_0} \{\widetilde{W}_0^*(N_*,l)\widetilde{W}_1(N_*,l)- \widetilde{C}_0^*(N_*,l) \widetilde{C}_1(N_*,l)\} dN_*+\text{c.c.} \label{inttype1}
\end{eqnarray}
is numerically computed for $N_0=20$.
The lower panel of the figure indeed shows that the jump vanishes at far field and this is due to the frozen property of the leading order inner wave as shown in the upper panel.
The proof showing $J(l)\equiv \lim_{N\rightarrow \infty} J_{N}(N,l)=0$ is given in Appendix A.2.

Thus the contribution to the left hand sides of (\ref{jumpp}) is purely from the outer wave terms.
\textcolor{black}{Those terms can easily be worked out by first transforming the wave into the pressure form (the relation similar to (\ref{incompinvwave2}) holds along the body-fitted coordinate, replacing $(y,z)$ by $(n,l)$) and then substituting the pressure expansion (\ref{incomplog}) with the phase shift given in (\ref{phase1}):
\begin{eqnarray}
~[\widetilde{\mathcal{V}}\widetilde{\mathcal{W}}^*-\widetilde{\mathcal{B}}\widetilde{\mathcal{C}}^*]^{+}_{-} +\text{c.c.}=
\left [\frac{q_nq_l^*+q_lq_n^*}{\alpha^2\Lambda}\right ]^{+}_{-} =
\frac{2\pi\text{sgn}(\lambda_0)}{\alpha^2|\lambda|}\Im (\widetilde{q}_0'^*\widetilde{q}_0''-\alpha^2\widetilde{q}_0'^*\widetilde{q}_0),
\end{eqnarray}
}
and $[|\widetilde{\mathcal{V}}|^2-|\widetilde{\mathcal{B}}|^2]^{+}_{-} +\text{c.c.}=[\widetilde{\mathcal{V}}\widetilde{\mathcal{C}}^*-\widetilde{\mathcal{B}}\widetilde{\mathcal{W}}^*]^{+}_{-} +\text{c.c.}=0$. 
Therefore, 
\begin{eqnarray}
~[\overline{Q}_0]^{\infty}_{-\infty}=[\overline{C}_{1N}]^{\infty}_{-\infty}=0,\qquad ~[\overline{W}_{1N}]^{\infty}_{-\infty}=\frac{\pi \Im(\alpha^{-2}\widetilde{Q}_0'^*\widetilde{Q}_0''-\widetilde{Q}_0'^*\widetilde{Q}_0
)}{\lambda_0|\lambda_u\pm \lambda_a|} .
\end{eqnarray}
Reverting back to the outer variables, the jump conditions at $y=f_{\pm}(z)$ can be found as
\begin{eqnarray}
~[\overline{q}]^{+}_{-}=[\{(\overline{b},\overline{c})\cdot \mathbf{e}_l\}_n]^{+}_{-}=0,\qquad
~[\{(\overline{v},\overline{w})\cdot \mathbf{e}_l\}_n]^{+}_{-}=\left . \frac{\pi \Im(\alpha^{-2}\widetilde{q}^*_l\widetilde{q}_{ll}-\widetilde{q}^*_l
\widetilde{q}
)}{U|U_n\pm \overline{a}_n|} \right |_{y=f_{\pm}}~~~\label{incompjump}
\end{eqnarray}
(and $[\{(\overline{v},\overline{w})\cdot \mathbf{e}_n\}_n]^+_-=[\overline{b}_n]^+_-=[\overline{c}_n]^+_-=0$ as shown above).
The wave equation (\ref{incompinvwave}), the vortex equations (\ref{outvortex}), and the above jump conditions form a closed set of equations. 
The choice of $\epsilon=1$ means that the wave Reynolds-Maxwell stress in the outer region (the terms proportional to $\epsilon^2$ in (\ref{epsfeedback})) is not negligible.
That asymptotic structure is quite different from the vortex/Rayleigh wave interaction where $\epsilon=\delta^{1/2}\ll 1$ should be chosen, and hence, the outer wave Reynolds stress is negligible (Hall \& Smith 1991). 
However, in the next section, we shall show that if the two resonant curves $y=f_{\pm}(z)$ are close enough, the nature of the singularity becomes similar to the Hall-Smith theory, and the purely hydrodynamic case can indeed be recovered at the vanishing magnetic field limit.

\vspace{5mm}
\subsection{Type 2: the connection formulae for the case $(f_+-f_-)\sim O(\delta)$}
\begin{figure}
\centering
\includegraphics[scale=0.4,clip]{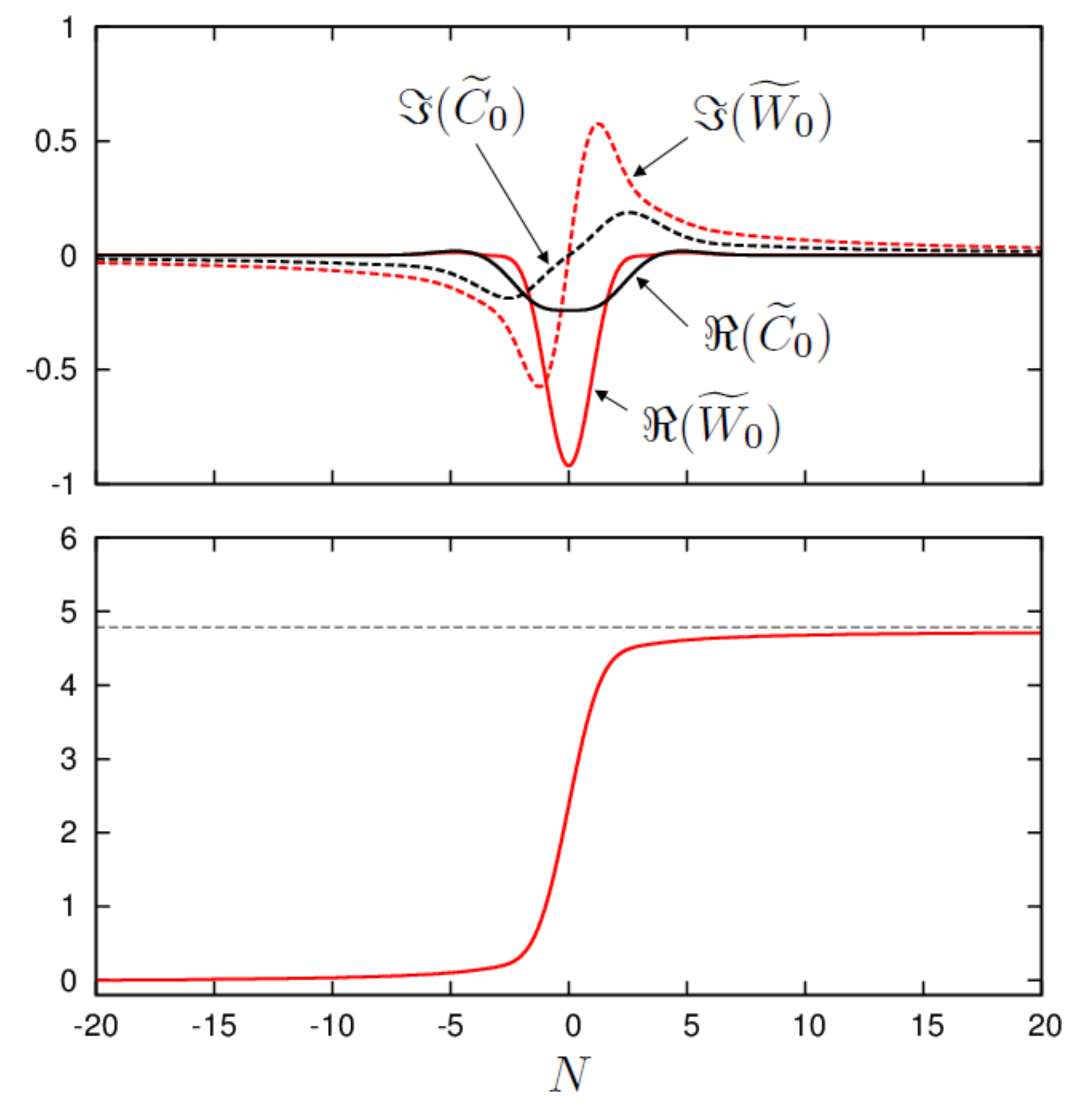}
\label{fig:bif}
\caption{
The similar plot as figure 3 but for type 2 interaction. The upper panel is the plot of the inner wave solution (\ref{waveWC}). The lower panel shows the corresponding wave forcing to the hydrodynamic roll component (\ref{jump22}). The limiting value calculated by (\ref{jump33}) is shown by the dashed line.
}
\end{figure}

When the distance between the two resonant curves $y=f_+(z)$ and $y=f_-(z)$ becomes comparable to \textcolor{black}{the dissipative layer thickness $\delta$ defined in (\ref{thickness})} or less we must consider another type of inner expansions; we refer to that interaction problem as type 2.

It is convenient to introduce the curve $y=f(z)$ near the two resonant curves, assuming that $(\overline{a}U_n-U\overline{a}_n)|_{n=0}$ is satisfied there.
In the vicinity of this curve, the fluid and magnetic streaks admit a small $n$ expansion
\begin{eqnarray}
U=\frac{\lambda_u}{\lambda_a}(\delta \lambda_0+\lambda_a n)+\cdots,\qquad
\overline{a}=(\delta \lambda_0+\lambda_a n)+\cdots,
\end{eqnarray}
where $\lambda_0,\lambda_u,\lambda_a$ are functions of the tangential coordinate $l$ attached to the curve $y=f(z)$.

If we define the new variable $\widehat{n}=n+\delta \lambda_0/\lambda_a$, then $\Lambda \sim O(\widehat{n}^2)$ when $\widehat{n}$ is small.
This means that the pressure Frobenius expansion for small $\widehat{n}$ can be written as
\begin{eqnarray}
\widetilde{q}=\widetilde{q}_0(l)+\widetilde{q}_2(l)\widehat{n}^2+\widetilde{q}_{3L}(l)\widehat{n}^3 (\ln|\widehat{n}|+i\Theta)+\widetilde{q}_3(l)\widehat{n}^3+\cdots.\label{logloglog}
\end{eqnarray}
\textcolor{black}{When we compare} this expansion and (\ref{incomplog}), we see that now the logarithmic singularity occurs at the higher order than type 1. 

Here, the coefficients $\widetilde{q}_0$ and $\widetilde{q}_3$ are the constants to be determined by the boundary conditions, and the substitution of the expansions $\Lambda=\lambda_1(l) \widehat{n}^2+\lambda_2(l) \widehat{n}^3+\cdots$ and (\ref{logloglog}) to (\ref{invwavebody}) yields
\begin{subequations}
\begin{eqnarray}
\widetilde{q}_2&=&\frac{1}{2}\left (\widetilde{q}_0''-\frac{\lambda_1'}{\lambda_1}\widetilde{q}_0'-\alpha^2 \widetilde{q}_0 \right ),\\
\widetilde{q}_{3L}&=&\frac{1}{3} \left \{
\left (\frac{2\lambda_2}{\lambda_1}-2\chi \right )\widetilde{q}_2
+2\chi \widetilde{q}_0'' +\chi' \widetilde{q}_0' +\frac{\lambda_1'}{\lambda_1}\left ( \frac{\lambda_2'}{\lambda_1'}-\frac{\lambda_2}{\lambda_1}-2\chi 
\right )\widetilde{q}_0'
\right \}.
\end{eqnarray}
\end{subequations}
The expansion (\ref{logloglog}) and the body-fitted coordinate expression of (\ref{incompinvwave2}) implies the small $\widehat{n}$ asymptotic behaviours
\begin{eqnarray}
\widetilde{u},\widetilde{a} \sim O(\widehat{n}^{-1}),\qquad
\widetilde{\mathcal{V}},\widetilde{\mathcal{B}} \sim O(\widehat{n}^0), \qquad 
\widetilde{\mathcal{W}},\widetilde{\mathcal{C}} \sim O(\widehat{n}^{-1}).
\end{eqnarray}

The above limiting forms suggest that the leading order inner expansions for $\widetilde{\mathcal{W}}$ and $\widetilde{\mathcal{C}}$ are unchanged from the previous case (\ref{innerwcexp}) to \textcolor{black}{the} leading order, but with the inner variable $N=\widehat{n}/\delta$, where the thickness remains $\delta=R^{-1/3}$. The normal component of the hydrodynamic wave equation again gives the pressure of the form $\widetilde{q}=\widetilde{Q}_0(l)+\cdots$, and then the $l$-component of the momentum and induction wave equations can be found as
\begin{subequations}
\begin{eqnarray}
i\alpha N (\lambda_u \widetilde{W}_0-\lambda_a \widetilde{C}_0)=-\widetilde{Q}_0'+\widetilde{W}_{0NN},\\
i\alpha N (\lambda_u \widetilde{C}_0-\lambda_a \widetilde{W}_0)=P_m^{-1}\widetilde{C}_{0NN},
\end{eqnarray}
\end{subequations}
to leading order. 
Multiplying $h P_m$ with a function $h(l)$ to the second equation and adding it to the first equation, we have
\begin{eqnarray}
i\alpha N \{ (\lambda_u-P_m h \lambda_a) \widetilde{W}_0+(P_m h \lambda_u-\lambda_a) \widetilde{C}_0 \}=-\widetilde{Q}_0'+(\widetilde{W}_{0NN}+h \widetilde{C}_{0NN}).\label{eqwc2}
\end{eqnarray}
If we choose $h(l)$ so that
\begin{eqnarray}
\frac{P_mh \lambda_u-\lambda_a}{\lambda_u-P_m h \lambda_a}=h,
\end{eqnarray}
namely
\begin{eqnarray}
h_{\pm}=\frac{1}{2}\left \{  -\frac{\lambda_u}{\lambda_a}(1-P_m^{-1})\pm \sqrt{\frac{\lambda_u^2}{\lambda_a^2}(1-P_m^{-1})^2+4P_m^{-1}} \right \},
\end{eqnarray}
then (\ref{eqwc2}) becomes
\begin{eqnarray}
\left ( \frac{\partial^2}{\partial \zeta^2}-\zeta  \right )(\widetilde{W}_0+h_{\pm}\widetilde{C}_0)=\frac{Q_0'}{(i\alpha |\lambda_{\pm}|)^{2/3}},
\end{eqnarray}
where 
\begin{eqnarray}
\lambda_{\pm}=\lambda_u -P_m h_{\pm}\lambda_a,\qquad \zeta_{\pm}=\text{sgn}(\lambda_{\pm})(i\alpha |\lambda_{\pm}|)^{1/3}N.
\end{eqnarray}
Therefore, the solution can again be solved in terms of the function $S$ defined in (\ref{defS}):
\begin{eqnarray}
\widetilde{W}_0+h_{\pm}\widetilde{C}_0=\frac{\widetilde{Q}_0'S(\zeta_{\pm})}{(i\alpha |\lambda_{\pm}|)^{2/3}}.
\end{eqnarray}
Here, we assumed $|\lambda_u|\neq|\lambda_a|$ for all $l$ (if that assumption is violated, then either $\lambda_+$ or $\lambda_-$ vanishes so the solution above breaks down).
Hence the $l$-component of the inner wave solutions are explicitly solved as
\begin{subequations} \label{waveWC}
\begin{eqnarray}
\widetilde{W}_0=\frac{\widetilde{Q}_0'}{h_+ -h_-}\left (\frac{h_+  S(\zeta_-)}{(i\alpha |\lambda_-|)^{2/3}} -\frac{h_-  S(\zeta_+)}{(i\alpha |\lambda_+|)^{2/3}} \right ),\\
\widetilde{C}_0=\frac{\widetilde{Q}_0'}{h_+ -h_-}\left ( \frac{S(\zeta_+)}{(i\alpha |\lambda_+|)^{2/3}}-\frac{S(\zeta_-)}{(i\alpha |\lambda_-|)^{2/3}} \right ).
\end{eqnarray}
\end{subequations}
The typical behaviour of the solutions is shown in the upper panel of figure 4.

Similar to the previous section, we can now derive the roll vorticity/current jump using the inner wave solutions.
\textcolor{black}{Like before,} $[\overline{\mathcal{V}}_{n}]^{+}_{-}=[\overline{\mathcal{B}}_{n}]^{+}_{-}=0$ from the solenoidal conditions; \textcolor{black}{additionally,} we need to examine the inner roll equations to find the jump occurring in the $l$-components.
Using the inner roll expansions shown in (\ref{exproll}) to the roll equations, the jump conditions can be found as
\begin{subequations}\label{jumpp2}
\begin{eqnarray}
\delta^{-1}\epsilon^2\chi \int^{\infty}_{-\infty} (|\widetilde{W}_0|^2-|\widetilde{C}_0|^2)dN +\text{c.c.}+\cdots=[\overline{Q}_{0}]^{\infty}_{-\infty},~~~\\
\delta^{-1}\epsilon^2\frac{\partial}{\partial l}\int^{\infty}_{-\infty} (|\widetilde{W}_0|^2-|\widetilde{C}_0|^2) dN +\text{c.c.}+\cdots=[\overline{W}_{1N}]^{\infty}_{-\infty},~~~\\
~\textcolor{black}{ \epsilon^2P_m\left \{ [\widetilde{\mathcal{V}}\widetilde{\mathcal{C}}^*-\widetilde{\mathcal{B}}\widetilde{\mathcal{W}}^*]^{+}_{-}+\chi \int^{\infty}_{-\infty}(\widetilde{B}_0\widetilde{W}_0^*-\widetilde{V}_0\widetilde{C}_0^*)dN \right \} }+\text{c.c.}+\cdots=[\overline{C}_{1N}]^{\infty}_{-\infty}.~~~\label{jumppc}
\end{eqnarray}
\end{subequations}
The above results analogue to (\ref{jumpp}) for type 1 but here $(|\widetilde{W}_0|^2-|\widetilde{C}_0|^2)$ does not vanish as the leading order inner magnetic wave is not frozen. 
Therefore, in order to balance the roll shear and the forcing terms from the waves, we must choose the smaller wave amplitude 
\begin{eqnarray}
\epsilon =\delta^{1/2}
\end{eqnarray}
than the previous case.

The numerical computation of 
\begin{eqnarray}
J_{N_0}(N,l)=\int^{N}_{-N_0} \{|\widetilde{W}_0(N_*,l)|^2-|\widetilde{C}_0(N_*,l)|^2\}dN_*+\text{c.c.}\label{jump22}
\end{eqnarray}
for $N_0=20$ shown in the lower panel of figure 4 suggests that there is indeed a contribution from the inner wave to the jump in the roll vorticity.
The limiting behaviour $J(l)\equiv \lim_{N\rightarrow \infty}J_{N}(N,l)$
can be found analytically (see Appendix A.3):
\begin{eqnarray}
J=\frac{2|\widetilde{Q}_0'|^2\textcolor{black}{G_0}(0)}{(\alpha|\lambda_+|)^{5/3}(h_+-h_-)^2}
\left \{
(h_+^2-1)r^{5/3}+ (h_-^2-1)+\frac{2^{4/3}(1-h_+h_-)r}{|1+\sigma r|^{1/3}}
\right \},\label{jump33}
\end{eqnarray}
where
\begin{eqnarray}
G_0(0)=\pi(2/3)^{2/3}\Gamma(1/3)\approx 6.4227\label{g0g0g0g0}
\end{eqnarray}
(the function $G_0(X)$ is to be defined in (\ref{defG})). 
The limiting value calculated by (\ref{jump33}) for the special case is indicated in the lower panel of figure 4 by the dashed line. 

Given $J$, the jump conditions are obtained as
\begin{eqnarray}
~[\overline{V}_{1N}]^{\infty}_{-\infty}=[\overline{B}_{1N}]^{\infty}_{-\infty}=[\overline{C}_{1N}]^{\infty}_{-\infty}=0,~~~~~~[\overline{Q}_0]^{\infty}_{-\infty}=\chi J,\qquad~[\overline{W}_{1N}]^{\infty}_{-\infty}=J'.~~\label{jumptype2}
\end{eqnarray}
The jump conditions rewritten in terms of the outer variables, the wave equation (\ref{incompinvwave}), and the vortex equations (\ref{outvortex}) constitute a closure: now, $\epsilon$ is chosen to be small so the terms proportional to $\epsilon^2$ is negligible in (\ref{epsfeedback}).  

To close this section, we shall make a few remarks. First, the analytic expression of $J$ is still valid for the limit of $P_m\rightarrow \infty$:
\begin{eqnarray}
J(l)=\frac{2|\widetilde{Q}_0'|^2G_0(0)}{(\alpha|\lambda_u|)^{5/3}}\frac{\lambda_u^2-\lambda_a^2}{\lambda_u^2}.
\end{eqnarray}
The limiting form of the vortex equations can be easily found by simply dropping the terms proportional to $P_m^{-1}$ from (\ref{outvortex}). 
\textcolor{black}{
Note that the jump in the leading order magnetic roll remains zero despite the factor $P_m$ in (\ref{jumppc}). This is because 
any non-zero jump should be produced by the resistive part of the wave, which is $O(P_m^{-1})$ times smaller than the leading order wave in the dissipative layer.
}

Second, 
if we take the hydrodynamic limit $\lambda_a \rightarrow 0$ in (\ref{jump33}),
the jump condition reduces to that for the vortex/Rayleigh wave interaction:
\begin{eqnarray}
J(l)=\frac{2|\widetilde{Q}_0'|^2G_0(0)}{(\alpha|\lambda_u|)^{5/3}}.
\end{eqnarray}
This is of course the expected result because in the absence of the magnetic effect, the entire system should become the reduced Navier-Stokes equations derived by Hall \& Smith (1991).

\subsection{The interaction diagram}
\begin{figure}
\centering
\includegraphics[scale=0.3,clip]{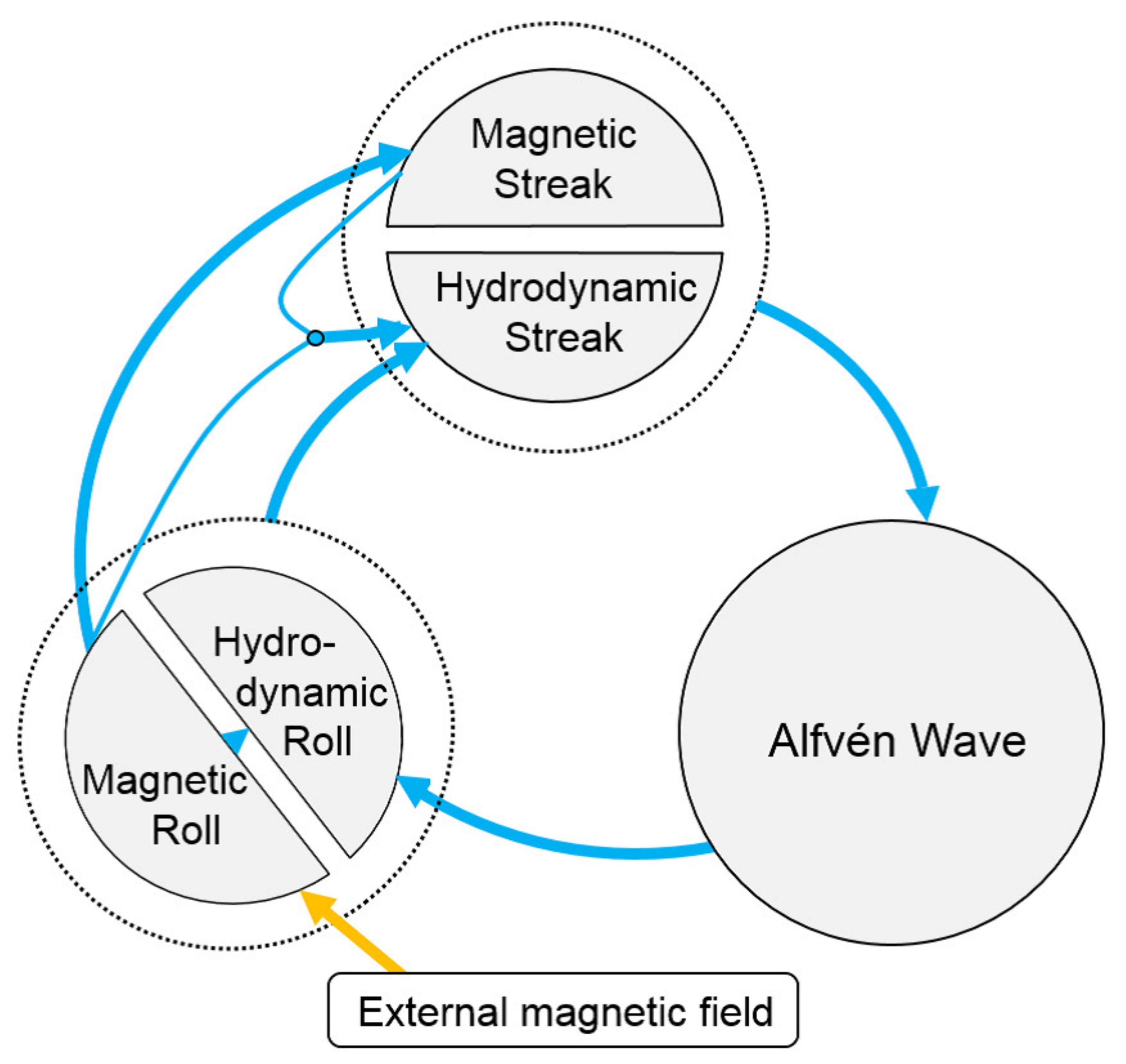}
\label{fig:bif}
\caption{
Illustration of the interaction diagram inferred by the vortex/Alfv\'en wave interaction theory. 
}
\end{figure}

Here, we shall clarify the physical driving mechanism of the vortex/Alfv\'en wave interaction states using the derived asymptotic closure.
We begin by reviewing the sustainment mechanism of the hydrodynamic roll-streak flow, for example, \textcolor{black}{studied by} Waleffe (1997).
Consider the vortex equations (\ref{invmo}) at finite Reynolds numbers. 
If we define the roll stream function $\psi$ so that $\overline{v}=\psi_z, \overline{w}=-\psi_y$, we have
the forced advection-diffusion equation for the hydrodynamic streak and the roll vorticity $\omega=-\triangle_2\psi$,
\begin{eqnarray}
P_m^{-1}(\psi_z\partial_y -\psi_y \partial_z)\overline{u}-\triangle_2 \overline{u}=\text{the forcing terms},\label{uueq} \\
P_m^{-1}(\psi_z\partial_y -\psi_y \partial_z)\omega-\triangle_2 \omega=\text{the forcing terms},\label{psieq}
\end{eqnarray}
where the forcing terms are from the right hand sides of (\ref{invmo}).
Then, multiplying $\overline{u}$ to (\ref{uueq}), $\psi$ to (\ref{psieq}), and integrating over the domain (here we denote the average over $[0,2\pi/\alpha]\times[-1,1]\times[0,2\pi/\beta]$ by $\langle ~\rangle$), we get
\begin{eqnarray}
\langle\overline{u}_y^2+\overline{u}_z^2 \rangle =\text{the terms associated with the forcing},\\
\langle \omega^2 \rangle =\text{the terms associated with the forcing},\label{intpsieq}
\end{eqnarray}
\textcolor{black}{which we integrate by parts and assuming that the boundary parts vanish. }
This is indeed the case if the domain is bounded by the periodic and no-slip boundaries. \textcolor{black}{The above equations imply that} $\omega=\overline{u}_y=\overline{u}_z=0$ everywhere, and we can further show that the perturbation is \textcolor{black}{actually} zero identically. Therefore, if there is no forcing term, then there is no non-trivial solution.

Likewise, we can define a `stream function' for the magnetic roll (namely toroidal magnetic potential) as $\overline{b}=\phi_z, \overline{c}=-\phi_y$. Then we can regard the magnetic streak and roll equations (\ref{invind}) as the forced advection-diffusion equations
\begin{eqnarray}
(\psi_z\partial_y -\psi_y \partial_z)\overline{a}-\triangle_2 \overline{a}=\text{the forcing terms},\label{advAA}\\
(\psi_z\partial_y -\psi_y \partial_z)\phi-\triangle_2 \phi=\text{the forcing terms}.\label{advPP}
\end{eqnarray}
Multiplying $\overline{a}$ to (\ref{advAA}), $\phi$ to (\ref{advPP}) and \textcolor{black}{integrating them} over the domain, we have
\begin{eqnarray}
\langle \overline{a}_y^2+\overline{a}_z^2 \rangle =\text{the terms associated with the forcing},\\
\langle \phi_y^2+\phi_z^2 \rangle=\text{the terms associated with the forcing}.\label{intphieq}
\end{eqnarray}
Here again, we assumed that the boundary terms vanish when they are integrated by parts: if the domain is surrounded by the periodic, perfectly insulating and/or perfectly conducting boundaries, the assumption is indeed valid.
Thus, similar to the hydrodynamic cases, there is no nontrivial solution if there is no forcing term.
This is the large radius limit manifestation of Cowling's anti-dynamo theorem (Cowling 1934) that motivated the physical interpretation of the MHD self-sustaining process by Rincon et al. (2007) and Rincon et al. (2008), where the Reynolds-Maxwell stresses and electromotive force by magneto-rotational instability waves are used to drive the roll components. More recently, Herreman (2018) related the sustained cycle and some critical layer-like structures appeared in the optimised perturbations of the induction equations at large magnetic Reynolds numbers.

Figure 5 shows the interaction diagram inspired by those works and, earlier Waleffe (1997), but here it is fully based on the asymptotically reduced equations for the vortex/Alfv\'en wave interaction.
The arrows in the figure show the interactions below:
\vspace{3mm}
\begin{itemize}
\item The arrow from the hydrodynamic roll to the hydrodynamic streak represents the lift-up term $U_b'\overline{v}$ in the hydrodynamic streak equation (see (\ref{epsfeedback})).
\item The arrow from the magnetic roll to the magnetic streak represents the term $U_b'\overline{b}$ in the magnetic streak equation (see (\ref{zerofeedback})). In the dynamo terminology this term corresponds to the omega effect.
\item The arrow from the magnetic roll and the magnetic streak to the hydrodynamic streak represents the roll-streak Lorentz force terms in the hydrodynamic streak equation. The small arrow from the magnetic roll to the hydrodynamic roll also represents the Lorentz force due to the self-interaction of the magnetic rolls.
\item Both the hydrodynamic and magnetic streaks create the Alfv\'en wave instability through (\ref{incompinvwave}), as shown by the arrow from the streak to the wave. 
\item The jump conditions (\ref{incompjump}) or (\ref{jumptype2}) produce the forcing to the hydrodynamic roll as indicated by the arrow from the wave to the hydrodynamic roll. In type 2, that jump is the sole mechanism of the feedback effect, whilst in type 1, it also occurs in the outer region. In the dynamo terminology, those feedbacks correspond to the alpha effect.
\end{itemize}
\vspace{3mm}
The most striking feature of this diagram is that there is no forcing mechanism to the magnetic roll components.
This means that the leading order part of the magnetic roll cannot be maintained without some external field to \textcolor{black}{support} it.
In the absence of the magnetic roll, the omega-effect is lost, so we have \textcolor{black}{a} small leading order magnetic streak. 
This means that the wave instability is now purely hydrodynamic; so we are left \textcolor{black}{merely} with the hydrodynamic self-sustained process with no magnetic effect.

\textcolor{black}{
We could add a small external magnetic field of $O(R_m^{-1})$ to drive 
the magnetic roll component. However, note that the generated much bigger magnetic streak of $O(1)$ when the $\overline{b}$ component is supported by the external field is not a dynamo in usual sense, because this is merely the amplification through the omega effect. 
}

\textcolor{black}{
Nevertheless, the vortex/Alfv\'en wave interaction states can be used to generate nontrivial amplification mechanisms of magnetic fields. 
For example, consider place Couette flow with a small external uniform spanwise magnetic field of $O(R_m^{-1})$ (or more generally, assume that it does not depend on $z$). In that case the generation of the $\overline{b}$ component is non-trivial so it is not clear if the omega effect can be driven. The flow remains always linearly stable. 
However, the nonlinear finite Reynolds number computation in \textcolor{black}{Deguchi (2019)} indeed found the non-trivial streamwise magnetic field amplification. 
The generation of that magnetic streak must be explained by the entire vortex/Alfv\'en wave interaction mechanism shown in the figure.
More precisely, the $\overline{b}$ component needed for the omega effect is induced by the hydrodynamic roll, which should be maintained by the three-dimensional Alfv\'en wave. 
}

\section{The self-sustained shear driven dynamo theory}
The result derived in the last section does not actually mean that dynamos are not possible in the limit of zero external magnetic fields. 
This is because the conclusion at the end of section 3 only inhibits the presence of the magnetic field whose size is appropriate to the leading order vortex/Alfv\'en interaction. In fact, \textcolor{black}{a dynamo} theory is possible for the purely shear driven cases if we choose slightly smaller \textcolor{black}{sized} leading order magnetic fields. The asymptotic theory here is \textcolor{black}{partially} motivated by the numerical finding of such dynamos to be shown in \textcolor{black}{Deguchi (2019).}

In order to formulate the asymptotic theory for such self-sustained shear driven dynamos, S$^3$ dynamos for short, we consider the leading order \textcolor{black}{of the} roll-streak-wave magnetic field \textcolor{black}{to be} smaller than the corresponding hydrodynamic parts by the factor of $\delta_m\ll1$. 
One may think the dynamo becomes merely kinematic in that case, but \textcolor{black}{later it turned} out that the magnetic wave is more amplified than the hydrodynamic wave around the resonant curve to drive nonlinear dynamo process. That amplification also allows the inner wave electromotive force to cause finite amount of jumps in the outer roll current so that any additional external force is no more necessary to drive the whole dynamo process as opposed to the previous section.

The outer expansions of the S$^3$ dynamos are
\begin{subequations}
\begin{eqnarray}
\left[ \begin{array}{c} u\\ v\\ w\\q \end{array} \right]=
\left[ \begin{array}{c} U_b+\overline{u}\\ R^{-1}\overline{v}\\ R^{-1}\overline{w}\\  R^{-2}\overline{q}  \end{array} \right]
+\epsilon R^{-1} \left \{ e^{i \alpha x}
\left[ \begin{array}{c} \widetilde{u}\\ \widetilde{v}\\ \widetilde{w}\\ \widetilde{q}\end{array} \right]
+\text{c.c.}
\right \},\\
\left[ \begin{array}{c} a\\b\\c \end{array} \right]=\delta_m
\left[ \begin{array}{c}   \overline{a}\\ R^{-1}\overline{b}\\ R^{-1}\overline{c} \end{array} \right]
+\epsilon \delta_m R^{-1} \left \{ e^{i \alpha x}
\left[ \begin{array}{c}  \widetilde{a}\\ \widetilde{b}\\ \widetilde{c}\\ \end{array} \right]
+\text{c.c.}
\right \}.\label{expansionS3mag}
\end{eqnarray}
\end{subequations}
Here and hereafter, we restrict our attention to $P_m=1$ for the sake of simplicity (otherwise the analysis become much more complicated, although the asymptotic closure can be found similar way). 
Under this outer scaling, we have the leading order outer roll-streak equations
\begin{subequations}\label{s3outer}
\begin{eqnarray}
~[(\overline{v}\partial_y +\overline{w}\partial_z)-(\partial_y^2+\partial_z^2)]
\left[ \begin{array}{c} \overline{u}\\ \overline{v}\\ \overline{w} \end{array} \right]
+\left[ \begin{array}{c} 0\\ \overline{q}_y\\ \overline{q}_z \end{array} \right]
=
 \left[ \begin{array}{c} -U_b'\overline{v} \\ 0\\ 0 \end{array} \right],~~~
\\
~[(\overline{v}\partial_y +\overline{w}\partial_z)-(\partial_y^2+\partial_z^2)]
\left[ \begin{array}{c} \overline{a}\\ \overline{b}\\ \overline{c} \end{array} \right]
-[\overline{b}\partial_y +\overline{c}\partial_z]
\left[ \begin{array}{c} 0\\ \overline{v}\\ \overline{w} \end{array} \right]=
\left[ \begin{array}{c} U_b' \overline{b}+(\overline{b}\partial_y +\overline{c}\partial_z)\overline{u}\\ 0 \\ 0\end{array} \right],~~~
\\
\overline{v}_y+\overline{w}_z=0, \qquad \overline{b}_y+\overline{c}_z=0, ~~~~~
\end{eqnarray}
and the leading order outer wave equations
\begin{eqnarray}
\left ( \frac{\widetilde{q}_y}{U^2} \right )_y
+\left ( \frac{\widetilde{q}_z}{U^2} \right )_z
-\alpha^2\frac{\widetilde{q}}{U^2}
=0, \qquad U=U_b+\overline{u}-s,\label{incompinvwaveS3}
\end{eqnarray}
\end{subequations}
which becomes singular when $U=0$. \textcolor{black}{The other wave components can be related to $\widetilde{q}$ through (\ref{incompinvwave2}).} 
The outer Lorentz force terms are completely neglected so no feedback from the magnetic field to the hydrodynamic field can be found in the above equations. 
However, as we shall see shortly, the feedback \textcolor{black}{does exist} within the dissipation layer occurring around the resonant curve. 

Let us write the location of the resonant curve as $y=f(z)$ and consider the body-fitted coordinate $(n,l)$ attached to that curve.
There is one resonant curve in the flow so the behaviour of the outer wave pressure is the same as type 2; see (\ref{logloglog}).
The key to the S$^3$ dynamo theory is the behaviours of the wave near the resonant curve
\begin{subequations}\label{singularS3}
\begin{eqnarray}
\widetilde{u}\sim O(n^{-1}), ~~~\widetilde{\mathcal{V}}\sim O(n^{0}), ~~~\widetilde{\mathcal{W}}\sim O(n^{-1}),\\
\widetilde{a}\sim O(n^{-2}),~~~\widetilde{\mathcal{B}}\sim O(n^{-1}),~~~\widetilde{\mathcal{C}}\sim O(n^{-2}),
\end{eqnarray}
\end{subequations}
inferred by the wave pressure expansion and the body-fitted coordinate version of (\ref{incompinvwave2}).
The behaviour (\ref{singularS3}) implies that near the resonant curve \textcolor{black}{the small outer wave magnetic field is more amplified than the hydrodynamic conterpart. }
Thus, if we choose
\begin{eqnarray}
\qquad \delta_m=\delta,\label{scaleDm}
\end{eqnarray}
\textcolor{black}{with $\delta$ defined at $(\ref{thickness})$}, then within the dissipative layer, the hydrodynamic and magnetic waves become comparable in size. 
In that case, the corresponding inner wave expansions can be found as
\begin{subequations}\label{expS3}
\begin{eqnarray}
\widetilde{q}=\widetilde{Q}_0(N,l)+\delta^2 \widetilde{Q}_2(N,l)+(\delta^3 \ln \delta) \widetilde{Q}_{3L}(N,l)+\delta^3 \widetilde{Q}_3(N,l)+\cdots,\\
\widetilde{u}=\delta^{-1}\widetilde{U}_0(N,l)+\cdots,\qquad
\widetilde{a}=\delta^{-2}\widetilde{A}_0(N,l)+\cdots,\\
\widetilde{\mathcal{V}}=\widetilde{V}_0(N,l)+\cdots,\qquad
\widetilde{\mathcal{B}}=\delta^{-1}\widetilde{B}_0(N,l)+\cdots,\\
\widetilde{\mathcal{W}}=\delta^{-1}\widetilde{W}_0(N,l)+\cdots,\qquad
\widetilde{\mathcal{C}}=\delta^{-2}\widetilde{C}_0(N,l)+\cdots,
\end{eqnarray}
\end{subequations}
where $N=n/\delta$ with the thickness $\delta=R^{-1/3}$, whilst the roll field expansions are unchanged from (\ref{exproll}).

As remarked earlier, the scaling (\ref{scaleDm}) is motivated so that both of the wave Reynolds and Maxwell stresses contribute to the vorticity jump in the roll field.
The leading order part of the $l$- and $n$-components of the hydrodynamic roll equations yield
\begin{subequations}\label{s3roll}
\begin{eqnarray}
\frac{\partial}{\partial l}\int^{\infty}_{-\infty}(|\widetilde{W}_0|^2-|\widetilde{C}_0|^2)dN+\text{c.c.}=[\overline{W}_{1N}]^{\infty}_{-\infty},\\
\chi \int^{\infty}_{-\infty}(|\widetilde{W}_0|^2-|\widetilde{C}_0|^2)dN+\text{c.c.}=[\overline{Q}_0]^{\infty}_{-\infty},
\end{eqnarray}
where we choose $\epsilon=\delta^{1/2}$ of type 2. 
As we shall see shortly, the left hand sides of the above equations are non-vanishing, so the hydrodynamic and magnetic fields are strongly coupled. 

On the other hand, for the inner magnetic roll equations, the balance of the magnetic roll stress and the wave electromotive force must be altered from type 2.
Now the magnetic roll size is $O(\delta)$ smaller from (\ref{expansionS3mag}) with (\ref{scaleDm}), but the inner magnetic wave size is unchanged from type 2 due to the singular behaviour (\ref{singularS3}). 
As a result, the $l$-component of the magnetic roll equations becomes
\begin{eqnarray}
\chi \int^{\infty}_{-\infty}(\widetilde{B}_0\widetilde{W}_0^*-\widetilde{V}_0\widetilde{C}_0^*)dN+\text{c.c.}=[\overline{C}_{1N}]^{\infty}_{-\infty}.\label{s3rollmag}
\end{eqnarray}
\end{subequations}
The above equation suggests that there is a finite current jump occurring in the magnetic roll component unlike the vortex/Alfv\'en interaction. Therefore, the dynamo can be driven purely by shear, without any external \textcolor{black}{force on} the magnetic roll component.

The derivation of the left hand sides in (\ref{s3roll}) is more complicated than the previous sections, although a first few steps are similar.
Upon making use of expansions (\ref{expS3}) to the hydrodynamic and magnetic wave equations and neglecting small terms, we have
\begin{subequations}
\begin{eqnarray}
i\alpha (\lambda N \widetilde{U}_0 -\gamma  \widetilde{A}_0)+\lambda \widetilde{V}_0+\lambda'N\widetilde{W}_0-\gamma '\widetilde{C}_0=-i\alpha \widetilde{Q}_0 +\widetilde{U}_{0NN},\label{s3hyx}\\
0=-\widetilde{Q}_{0N},\label{s3hyn}\\
i\alpha (\lambda N \widetilde{W}_0 -\gamma \widetilde{C}_0)=-\widetilde{Q}_{0l}+\widetilde{W}_{0NN},\label{s3hyl}\\
i\alpha (\lambda N \widetilde{A}_0 -\gamma  \widetilde{U}_0)-\lambda \widetilde{B}_0-\lambda'N
\widetilde{C}_0+\gamma '\widetilde{W}_0=\widetilde{A}_{0NN},\label{s3mgx}\\
i\alpha (\lambda N \widetilde{B}_0 -\gamma  \widetilde{V}_0)=\widetilde{B}_{0NN},\label{s3mgn}\\
i\alpha (\lambda N \widetilde{C}_0 -\gamma  \widetilde{W}_0)=\widetilde{C}_{0NN},\label{s3mgl}
\end{eqnarray}
where, we used the Taylor expansions of the hydrodynamic streak $U=\lambda(l) n+\cdots$ and the magnetic streak $\overline{a}=\gamma(l)+\cdots$ around $n=0$.
\textcolor{black}{Additionally,} from the solenoidal conditions,
\begin{eqnarray}
i\alpha \widetilde{U}_0+\widetilde{V}_{0N}+\widetilde{W}_{0l}=0,\label{s3hyc}\\
i\alpha \widetilde{A}_0+\widetilde{B}_{0N}+\widetilde{C}_{0l}=0.\label{s3mgc}
\end{eqnarray}
\end{subequations}
Those equations must be solved subject to the matching conditions at the far-field $|N|\rightarrow \infty$, namely $\widetilde{Q}_0\rightarrow \widetilde{q}_0$ and
\begin{subequations}
\begin{eqnarray}
\widetilde{U}_0\rightarrow -\frac{\widetilde{q}_0''-\frac{\lambda'}{\lambda}\widetilde{q}_0'}{\lambda \alpha^2}\frac{1}{N},~~~
\widetilde{V}_0\rightarrow -\frac{\widetilde{q}_0''-\frac{2\lambda'}{\lambda}\widetilde{q}_0'-\alpha^2\widetilde{q}_0}{i\alpha \lambda},~~~
\widetilde{W}_0\rightarrow -\frac{\widetilde{q}_0'}{i\alpha \lambda}\frac{1}{N},~~~~\\
\widetilde{A}_0\rightarrow -\frac{\frac{\gamma'}{\gamma}\widetilde{q}'_0+\alpha^2\widetilde{q}_0}{\lambda \alpha^2}\frac{N_0}{N^2},~
\widetilde{B}_0\rightarrow -\frac{\widetilde{q}_0''-\frac{2\lambda'}{\lambda}\widetilde{q}_0'-\alpha^2\widetilde{q}_0}{i\alpha \lambda}\frac{N_0}{N},~
\widetilde{C}_0\rightarrow  -\frac{\widetilde{q}_0'}{i\alpha \lambda}\frac{N_0}{N^2},~~~~
\end{eqnarray}
\end{subequations}
where $N_0(l)=\gamma/\lambda$.

From (\ref{s3hyn}), the pressure $\widetilde{Q}_0$ is a function of $l$ only and equals to $\widetilde{q}_0$. 
Then we can solve for $\widetilde{W}_0, \widetilde{C}_0$ using the $l$-components of the wave equations (\ref{s3hyl}), (\ref{s3mgl}). Those equations can be simplified as
\begin{eqnarray}
\left (\frac{\partial^2}{\partial \zeta_{\mp}^2}-\zeta_{\mp}  \right )(\widetilde{W}_0\pm \widetilde{C}_0)=\frac{\widetilde{Q}_0'}{(i\alpha |\lambda|)^{2/3}},\label{simpWC}
\end{eqnarray}
using the new variables
\begin{eqnarray}
\zeta_{\pm}=\text{sgn}(\lambda)(i\alpha|\lambda|)^{1/3}(N\pm N_0).
\end{eqnarray}
The solutions of (\ref{simpWC}) can be found in terms of the function $S$ defined in (\ref{defS}), and we have
\begin{eqnarray}
\widetilde{W}_0=\frac{\widetilde{Q}_0'}{(i\alpha |\lambda|)^{2/3}}\frac{S(\zeta_{-})+S(\zeta_{+})}{2},\qquad
\widetilde{C}_0=\frac{\widetilde{Q}_0'}{(i\alpha |\lambda|)^{2/3}}\frac{S(\zeta_{-})-S(\zeta_{+})}{2}.\label{solutionWC}
\end{eqnarray}

Next from (\ref{s3hyx}), (\ref{s3mgx}), (\ref{s3mgn}) and (\ref{solutionWC}) we have the equations
\begin{eqnarray}
\left (\frac{\partial^2}{\partial \zeta_{\mp}^2}-\zeta_{\mp}  \right )\left (\widetilde{U}_0+\frac{\lambda}{i\alpha \gamma}\widetilde{B}_0\pm \widetilde{A}_0 \right )
=\frac{(\lambda'N\pm \gamma')\widetilde{Q}_0'}{(i\alpha |\lambda|)^{4/3}}S(\zeta_{\pm})+\frac{i\alpha \widetilde{Q}_0}{(i\alpha |\lambda|)^{2/3}},
\end{eqnarray}
which give the solutions
\begin{eqnarray}
\widetilde{U}_0+\frac{\lambda}{i\alpha \gamma}\widetilde{B}_0 &=&\left (i\alpha \widetilde{Q}_0-\frac{\lambda'\widetilde{Q}_0'}{i\alpha \lambda}\right )\frac{S(\zeta_+)+S(\zeta_-)}{2(i\alpha |\lambda|)^{2/3}}+\frac{\lambda'\widetilde{Q}_0'}{(i\alpha \lambda)N_0}\frac{\zeta_+S(\zeta_+)-\zeta_-S(\zeta_-)}{2(i\alpha |\lambda|)},~~~~~\\
\widetilde{A}_0&=&\left (\frac{\gamma' \widetilde{Q}_0'}{i\alpha \gamma}-i\alpha \widetilde{Q}_0 \right )\frac{S(\zeta_+)-S(\zeta_-)}{2(i\alpha |\lambda|)^{2/3}}.\label{solutionA}
\end{eqnarray}

Now $\widetilde{A}_0,\widetilde{C}_0$ given in (\ref{solutionWC}), (\ref{solutionA}) can be used to integrate the magnetic solenoidal condition (\ref{s3mgc}) with respect to $N$. 
Selecting the integration constant to satisfy the matching condition, the normal component of the magnetic field can be found as
\begin{eqnarray}
\widetilde{B}_0=-\widetilde{Q}_{00} \frac{\kappa(\zeta_-)-\kappa(\zeta_+)}{2(i\alpha |\lambda|)}-\frac{\lambda'\widetilde{Q}_0'}{3\lambda}\frac{\zeta_-S(\zeta_-)-\zeta_+S(\zeta_+)}{2(i\alpha |\lambda|)}+\widetilde{Q}_0'N_0'\frac{S(\zeta_-)+S(\zeta_+)}{2(i\alpha |\lambda|)^{2/3}},~~~
\end{eqnarray}
where the function $\kappa$ is defined in (\ref{defkappa}).

Finally, the last component $\widetilde{V}_0$ matching the outer solution can be found by integrating the continuity (\ref{s3hyc}). 
\begin{eqnarray}
\widetilde{V}_0&=&-\widetilde{Q}_{00} \frac{\kappa(\zeta_-)+\kappa(\zeta_+)}{2(i\alpha |\lambda|)}-\frac{\lambda'\widetilde{Q}_0'}{3\lambda}\frac{\zeta_-S(\zeta_-)+\zeta_+S(\zeta_+)}{2(i\alpha |\lambda|)}+\widetilde{Q}_0'N_0'\frac{S(\zeta_-)-S(\zeta_+)}{2(i\alpha |\lambda|)^{2/3}}\nonumber\\
&&-\frac{\widetilde{Q}_{00}}{N_0}\frac{\zeta_-\kappa(\zeta_-)+\zeta_+\kappa(\zeta_+)}{2(i\alpha |\lambda|)^{4/3}}
+\frac{1}{N_0}\left (\widetilde{Q}_{00}+\frac{2\lambda'\widetilde{Q}_0'}{3\lambda} \right)\frac{S'(\zeta_-)-S'(\zeta_+)}{2(i\alpha |\lambda|)^{4/3}}\nonumber \\
&&-\frac{1}{i\alpha |\lambda|}\left (2\widetilde{Q}_{00}-\frac{4\lambda'\widetilde{Q}_0'}{3\lambda}+\frac{\gamma'\widetilde{Q}_0'}{\gamma}\right ).
\end{eqnarray}

Using the above wave solutions to (\ref{s3roll}) and doing some algebra (see Appendix A.4), the jumps in terms of the outer variables can be found as
\begin{subequations}\label{s3jumpfinal}
\begin{eqnarray}
~[\overline{q}]^{0_+}_{n=0_-}&=&\chi \left . \left (\frac{2|\widetilde{q}_l |^2G_0}{(\alpha |U_n|)^{5/3}}\right )\right |_{y=f}, \\
~[\{(\overline{v},\overline{w})\cdot \mathbf{e}_l\}_n]^{0_+}_{n=0_-}&=&\left . \left (\frac{2|\widetilde{q}_l |^2G_0}{(\alpha |U_n|)^{5/3}}\right )_l\right |_{y=f},\\
~[\{(\overline{b},\overline{c})\cdot \mathbf{e}_l\}_n]^{0_+}_{n=0_-}&=&\chi \frac{(\overline{a}/U_n)}{(\alpha |U_n|)^{5/3} }\left . \left (
\left \{ |\widetilde{q}_l|^2-\alpha^2 |\widetilde{q}|^2 \right \}_lG_1 \vbox to 14pt{} \right. \right. \nonumber \\
&&~~~~~\left.  \left .+2|\widetilde{q}_l|^2\left \{ \frac{(\overline{a}|U_n|^{-2/3})_l}{\overline{a}|U_n|^{-2/3}} G_0-\frac{(\overline{a}|U_n|^{2/3})_l}{\overline{a}|U_n|^{2/3}} G_1\right \} \right )\right |_{y=f},~~~~
\end{eqnarray}
\end{subequations}
where
\begin{eqnarray}
G_0(X)=2\pi\int^{\infty}_0\cos(2Xt)e^{-\frac{2}{3}t^3}dt,~~~~
G_1(X)=2\pi\int^{\infty}_0\frac{1-\cos(2Xt)}{X^2}te^{-\frac{2}{3}t^3}dt,\label{defG}
\end{eqnarray}
are evaluated at $X=(\alpha |U_n|)^{1/3}(\overline{a}/U_n)$. 
The numerically evaluated those special functions can be found in figure 6.

\begin{figure}
\centering
\includegraphics[scale=1.]{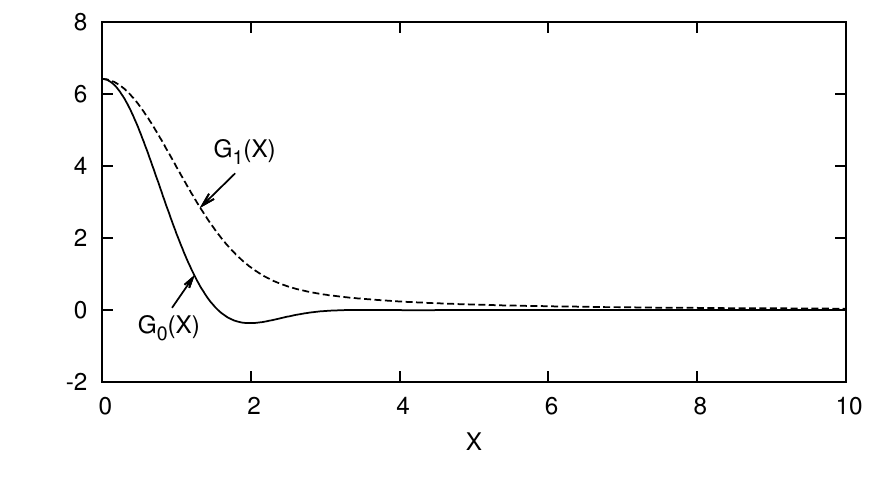} 
\caption{The functions $G_0$ and $G_1$ computed numerically. Note that these are even functions.}
\label{fig:longwave}
\end{figure}

\begin{figure}
\centering
\includegraphics[scale=0.3,clip]{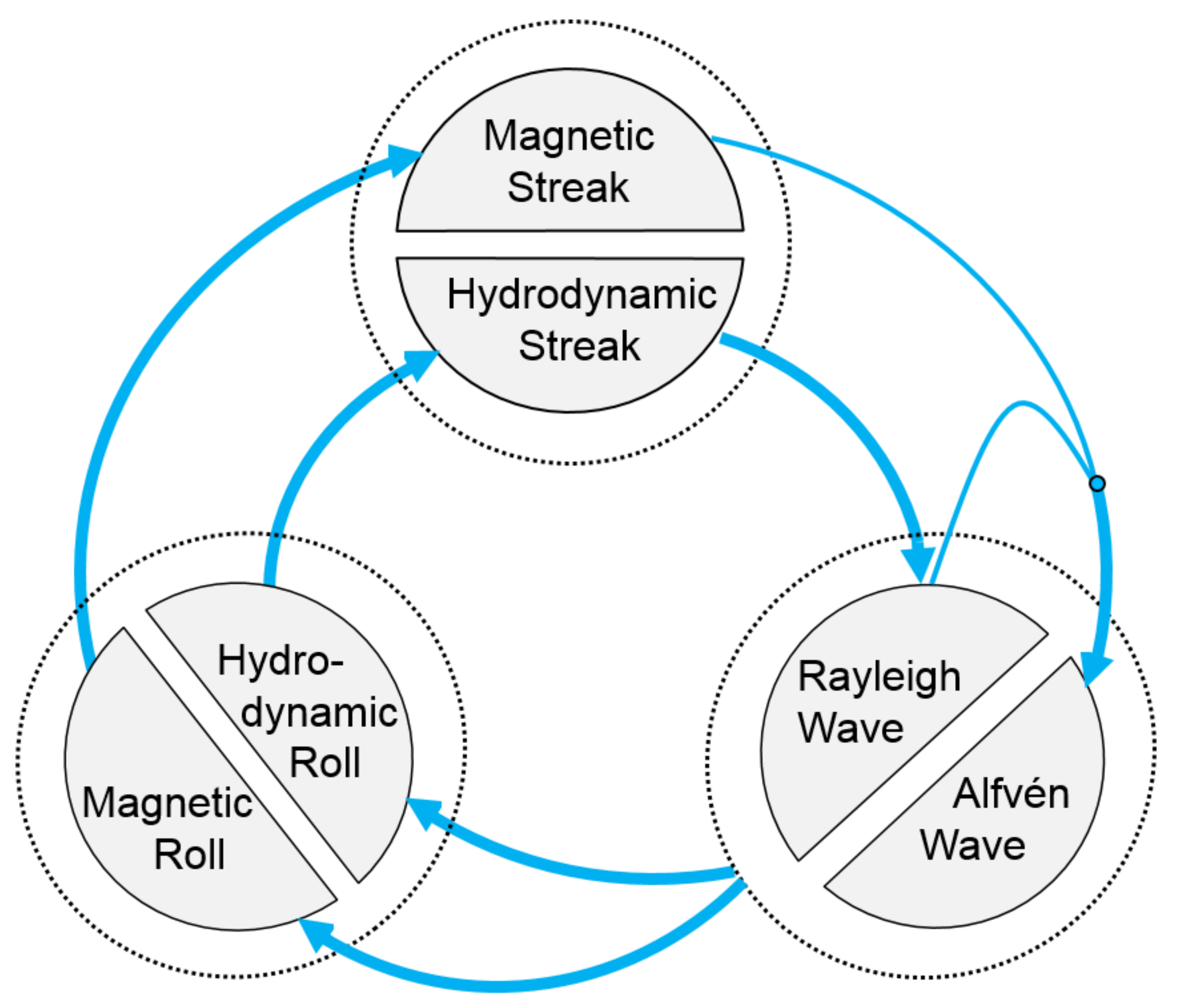}
\label{fig:bif}
\caption{
Illustration of the interaction diagram inferred by the S$^3$ dynamo theory. 
}
\end{figure}

The interaction diagram in figure 7 is formulated using the asymptotic closure for the S$^3$ dynamo, namely (\ref{s3outer}) and (\ref{s3jumpfinal}). 
There is no arrow from the magnetic roll/streak to the hydrodynamic roll/streak because there are no Lorentz force terms in the outer equations. 
The instability wave is merely Rayleigh wave to leading order, because it is purely driven by the hydrodynamic streak.
However, it interacts with the magnetic streak to produce Alfv\'en wave at \textcolor{black}{a} higher order. 
The Rayleigh and Alfv\'en waves amplified within the dissipative layer induce the feedback effects to both the hydrodynamic and magnetic roll components through the jumps, originated from the wave Reynolds-Maxwell stress and the wave electromotive force.

\section{Conclusion and discussion}

We have derived two three-dimensional nonlinear \textcolor{black}{MHD} theories using large Reynolds number matched asymptotic expansion of incompressible viscous resistive MHD equations. 
The closures at the asymptotic limit are found by considering the interaction between the roll, streak, and wave, defined both for the velocity and magnetic fields.
\textcolor{black}{The magnetic Prandtl number is assumed to be $O(R^0)$ or larger, except for Appendix B.}

The \textcolor{black}{primarily concerned} asymptotic \textcolor{black}{MHD} state is derived by considering similar roll-streak scaling as the usual vortex/wave interaction for both hydrodynamic and magnetic components.
Therefore the theory could be viewed as the third type of interaction, vortex/Alfv\'en wave interaction say, followed by the vortex/Tollmien-Schlichting wave interaction (\textcolor{black}{Hall \& Smith 1990; Bennett et al. 1991}; Dempsey et al. 2016) and the vortex/Rayleigh wave interaction (Hall \& Smith 1991; Hall \& Sherwin 2010).
The Alfv\'en wave is generated by the instability of the mean streamwise hydrodynamic and magnetic fields of $O(R^0)$ (namely, streaks). 
In the majority of the flow, the wave part satisfies the inviscid equation to leading order; \textcolor{black}{so} the cross-stream components of the magnetic wave field are frozen to those of the hydrodynamic wave field; see (\ref{frozen}). 

The wave amplitude must be found by the analysis within the dissipative layer of thickness $\delta=R^{-1/3}$ surrounding the Alfv\'en resonant curves \textcolor{black}{that} occur whenever the modulus of the doppler-shifted hydrodynamic streak and the modulus of the magnetic streak coincides.
As well-established in the resonant absorption theory, the wave there is amplified in a singular manner, although here, the singularity occurs on a curve rather than a point. 
The strong vortex and current sheets \textcolor{black}{that} emerged there strongly force the hydrodynamic roll components via the wave Reynolds and Maxwell stresses. The resultant vorticity jumps in the outer roll solutions can be found using the curved coordinate. 

If the two resonant curves are well separated, the leading order flow within the dissipative layer is still frozen because of the lack of the diffusivity at this order. 
The frozen effect cancels the leading order Reynolds and Maxwell stresses so that the feedback effect to the roll components becomes weaker than the hydrodynamic case. As a result, we must choose larger wave amplitude, which makes the outer wave Reynolds and Maxwell stress terms important as well. 
The smooth analytic solution within the dissipative layer, similar to the two-dimensional cases (Sakurai et al. 1991; Goossens et al. 1992; Goossens et al. 1995; Erd\'elyi et al. 1995; Eld\'elyi 1997), can be used to find how to connect the outer roll solutions across the resonant curves. 
\textcolor{black}{The scaling of the roll and streak are $O(R^{-1}P_m^{-1})$ and $O(1)$ respectively, which are typical for the viscous-resistive roll-streak field. 
Amplitude of the inviscid wave is $O(R^{-1}P_m^{-1/2})$ in the majority of the flow, but it is amplified within the dissipative layer of thickness $O(R^{-1/3})$ to be larger $O(R^{-2/3}P_m^{-1/2})$.}

The aforementioned scenario, called type 1, breaks down when the distance between the two resonant curves become less than the dissipative layer thickness. In that case, called type 2, the two dissipative layers \textcolor{black}{merge} and the nature of the singularity there becomes similar to the vortex/Rayleigh wave interaction.
The cancellation of the inner Reynolds and Maxwell stresses occurs \textcolor{black}{any more} because the leading order inner wave is dissipative. This means that the wave amplitude should be kept the same size as the hydrodynamic case. The analytic form of the jumps for type 2 can also be found, and it is shown that at the limit of vanishing magnetic effects, the jump conditions derived in Hall \& Smith (1991) can be recovered.
\textcolor{black}{The roll-streak scaling is unchanged from type 1, but for type 2 we must choose smaller outer wave amplitude of $O(R^{-7/6}P_m^{-1/2})$, which becomes larger $O(R^{-5/6}P_m^{-1/2})$ within the dissipative layer.}

For both types, the wave has an ability to modify the background flow to leading order, and hence, the nonlinear MHD \textcolor{black}{states} can be realised without any linear instability of the base flow. 
However, \textcolor{black}{an important caveat has been} found from the interaction diagram based on the asymptotic closure, pictured in figure 5;  the leading order magnetic fields of the vortex/Alfv\'en wave interaction states cannot be sustained without a weak external magnetic field forcing the magnetic roll components. 
The \textcolor{black}{nonlinear states nonetheless can drive} \textcolor{black}{a nontrivial magnetic field amplification mechanism} since if, for example, a spanwise uniform magnetic field of $O(R_m^{-1})$ is applied, they generate much bigger streamwise magnetic field of $O(1)$; \textcolor{black}{see the comments at the end of section 3.} 

Despite the caveat, for some cases, the dynamos of slightly different structures can be driven even when the external magnetic field is switched off, as \textcolor{black}{shown by the rather surprising numerical results in the companion paper} \textcolor{black}{Deguchi (2019).} Section 4 successfully formulated the asymptotic theory for that purely shear driven dynamos (the S$^3$ dynamos), revealing that the magnetic field generation indeed survives at the large Reynolds number limit.
\textcolor{black}{The key to the $S^3$ dynamo theory is that we have to take the different scaling for the hydrodynamic and magnetic components.}
\textcolor{black}{
The outer Alfv\'en wave of $O(R^{-3/2}P_m^{-1/2})$ is the secondary effect produced \textcolor{black}{as a result of} the interaction \textcolor{black}{between} the leading order outer Rayleigh wave of $O(R^{-7/6}P_m^{-1/2})$ and the magnetic streak of $O(R^{-1/3})$. 
The wave resonance, which is now found at the classical critical layer position, amplifies both the hydrodynamic and magnetic waves, to be the same size of $O(R^{-5/6}P_m^{-1/2})$}.
Therefore, the inner wave electromotive force is no more negligible, and hence, 
\textcolor{black}{both of the waves now have} an ability to drive both \textcolor{black}{the} hydrodynamic roll components \textcolor{black}{of $O(R^{-1}P_m^{-1})$ and \textcolor{black}{the} magnetic roll components of $O(R^{-4/3}P_m^{-1})$} to sustain the entire dynamo mechanism without any external magnetic field. 
The mean-flow modulation mechanisms, which in turn occur \textcolor{black}{as a result of} the lift-up and omega effects, are indispensable \textcolor{black}{for} generating the magnetic waves, 
\textcolor{black}{in view of} Zel'dovich's anti-dynamo theorem (Zel'dovich 1957).

\textcolor{black}{Those above travelling wave theories could be extended to describe more general time dependent solutions. However the formulation becomes much more complicated (see Deguchi \& Hall (2016) for purely hydrodynamic cases) and thus that possibility is omitted in this paper.
When the wave is not neutral, Ruderman et al. (1995) showed that the motion in the dissipative layer is very much different, although it turned out that the jump condition is not affected by the account of non-stationarity in the dissipative layer. }

Finally we briefly comment on the effect of the system rotation to the theory. 
When that effect exists, we must add the Coriolis term 
\begin{eqnarray}
\Omega \left[ \begin{array}{c} -v\\ u\\ 0 \end{array} \right]
\end{eqnarray}
to the left hand side of the momentum equations (\ref{govmomentum}).
Here, $\Omega$ is the non-dimensionalised rotation parameter, and the shear-Coriolis instability exists for $\Omega\in (0,1)$ according to Rayleigh's stability criterion. 
The Coriolis term brings a new interaction term from the streak to the roll.
We have assumed throughout the paper that the roll components are much smaller than the streak components in order to balance the viscous and convective effects in the roll-streak equations.
In order to further balance the new interaction term under this roll-streak scaling, we have to choose $\Omega \sim O(R_m^{-2})\ll 1$. 
This means that the Keplerian rotation of $\Omega=4/3$ destroys the assumed asymptotic balance of the asymptotic theories in this paper, and this may be the reason why it is so difficult to maintain nonlinear dynamo states at large Reynolds number regime (Rincon et al. (2007), Rincon et al. (2008), Riols et al. (2013)).

We further \textcolor{black}{remarked} here that the recent work by Deguchi (2017) nonetheless suggests the presence of the Keplerian asymptotic dynamo states if the vortices have short enough wavelengths. 
In the latter work, it was shown for stably stratified flow that small vortices at Kolmogorov microscale, found by Deguchi (2015), can persist for $O(1)$ Richardson number where the vortex/wave interaction states are impossible. 
Applying the analogy between the stratified and rotating flows, which has long been recognised (Veronis 1970), we arrive at the conjecture that the short wavelength asymptotic dynamo states might survive even when $\Omega \sim O(1)$. 

\textcolor{black}{
This work was supported by Australian Research Council Discovery Early Career Researcher Award DE170100171. The insightful comments of the referees should be gratefully acknowledged.}

\appendix

\section{Useful formulas}
\subsection{Derivation of the logarithmic jump}
The limiting form of the function $\kappa$ defined in (\ref{defkappa}) can be found as follows.
If we write $\zeta=i^{1/3}N_*$ and $\sigma=\text{sgn}(\lambda \lambda_0)$, then
\begin{eqnarray}
-\kappa&=&\int^{\infty}_0 \frac{1-e^{-i\sigma N_* t}}{t}e^{-\frac{t^3}{3}}dt\nonumber \\
&=&
\int^{\infty}_0 \frac{1-e^{-t^3/3}}{t}e^{-it\sigma N_*}dt+
\int^{\infty}_0 \frac{e^{-t^3/3}-\cos t}{t} dt \nonumber \\
&&+\int^{\infty}_0 \frac{\cos t -\cos (t N_*)}{t} dt+i\int^{\infty}_0 \frac{ \sin(\sigma t N_*)}{t} dt.
\end{eqnarray}
In the right hand side, 
the first term becomes small for large $|N_*|$ (integrate by parts), 
the second term is a constant, 
the third term is $\ln |N_*|$ (use cosine integral), 
and the fourth term is $\frac{i\pi}{2}\text{sgn}(\sigma N_*)$ (use Dirichlet integral). 

\subsection{Calculation of the limiting function $J$ for type 1}
Here we show $J(l)\equiv \lim_{N\rightarrow \infty} J_{N}(N,l)=0$ for the function defined in (\ref{inttype1}).
Firstly we note $(\widetilde{W}_0^*\widetilde{W}_1-\widetilde{C}_0^*\widetilde{C}_1)=\widetilde{W}_0^*(\widetilde{W}_1\pm \widetilde{C}_1)$ from (\ref{infrozen}) and further using (\ref{subwww1})--(\ref{subwww2})
\begin{eqnarray}
\int^{\infty}_{-\infty}(\widetilde{W}_0^*\widetilde{W}_1-\widetilde{C}_0^*\widetilde{C}_1) dN+\text{c.c.} 
=
\frac{i}{2\alpha\lambda_0}\int^{\infty}_{-\infty} \widetilde{W}_0^*\{\widetilde{Q}_{0}'+(P_m^{-1}-1)\widetilde{W}_{0NN}\} dN+\text{c.c.}~~~~
\label{wcwc}
\end{eqnarray}
Then from (\ref{incompwsol}) we can write $\widetilde{W}_0=\widetilde{Q}_0'(W_R+iW_I)$, where the real functions $W_R$ and $W_I$ have a symmetry
\begin{eqnarray}
W_R(-N,l)=W_R(N,l),\qquad W_I(-N,l)=-W_I(N,l),\label{symw}
\end{eqnarray}
see the upper panel of figure 3.
Hence the right hand side of (\ref{wcwc}) should vanish because the symmetry (\ref{symw}) implies that the integral is a purely imaginary function plus its complex conjugate.

\subsection{Calculation of the limiting function $J$ for type 2}
The limiting form of (\ref{jump22}) can be found as follows. First we note
\begin{eqnarray}
J&\equiv &\lim_{N\rightarrow \infty}J_{N}
(N,l)=\int^{\infty}_{-\infty} (|\widetilde{W}_0|^2-|\widetilde{C}_0|^2)dN+\text{c.c.} \nonumber \\
&=&\frac{2|\widetilde{Q}_0'|^2}{(h_+-h_-)^2} \left \{
(h_-^2-1)\int^{\infty}_{-\infty} \left |\frac{S(\zeta_+)}{(i\alpha |\lambda_+|)^{2/3}} \right |^2dN+(h_+^2-1)\int^{\infty}_{-\infty} \left |\frac{S(\zeta_-)}{(i\alpha |\lambda_-|)^{2/3}} \right |^2dN \right. \nonumber \\
&&\left . +(1-h_+h_-)\int^{\infty}_{-\infty} \left (\left (\frac{S(\zeta_-)}{(i\alpha|\lambda_-|)^{2/3}} \right)
\left (\frac{S(\zeta_+)}{(i\alpha|\lambda_+|)^{2/3}} \right)^*+\text{c.c.} \right ) dN \right \}. \label{A4}
\end{eqnarray}
If we write
\begin{eqnarray}
r=\frac{|\lambda_+|}{|\lambda_-|},\qquad
\sigma=\text{sgn}(\lambda_+\lambda_-)=\text{sgn}(\lambda_u^2-\lambda_a^2)
\end{eqnarray}
then we can compute
\begin{eqnarray}
&&\int^{\infty}_{-\infty}\left (\frac{S(\zeta_-)}{(i\alpha|\lambda_-|)^{2/3}} \right)
\left (\frac{S(\zeta_+)}{(i\alpha|\lambda_+|)^{2/3}} \right)^*dN \nonumber \\
&=&\frac{1}{(\alpha^2|\lambda_-||\lambda_+|)^{2/3}}
\int^{\infty}_{-\infty}  \int^{\infty}_0  \int^{\infty}_0 e^{-\frac{t^3+\tau^3}{3}}e^{i\text{sgn}(\lambda_-)(\alpha|\lambda_-|)^{1/3}\{\sigma r^{1/3} t-\tau\}N} \, dt d\tau dN \nonumber \\
&=&\frac{1}{(\alpha^2|\lambda_-||\lambda_+|)^{2/3}} \frac{2\pi}{(\alpha |\lambda_-|)^{1/3}}\int^{\infty}_0  \int^{\infty}_0 e^{-\frac{t^3+\tau^3}{3}}\widehat{\delta}  (\sigma r^{1/3} t- \tau )\, dt d\tau\nonumber \\
&=& \frac{1}{(\alpha^2|\lambda_-||\lambda_+|)^{2/3}} \frac{2\pi}{(\alpha |\lambda_-|)^{1/3}}\int^{\infty}_0  \int^{\infty}_0 e^{-(1+\sigma r)\frac{\tau^3}{3}}d\tau\nonumber \\
&=& \frac{rG_0(0)}{(\alpha |\lambda_+|)^{5/3}} \left (\frac{2}{1+\sigma r} \right )^{1/3},\label{formulaA6}
\end{eqnarray}
where we assumed $r<1$ (if $r>1$ the factor $(1+\sigma r)$ must be swapped by $(\sigma +r)$ in the final result) and the coefficient $G_0(0)$ is given in (\ref{g0g0g0g0}).
Here we have used the usual integral expressions of the Dirac delta function and the gamma function
\begin{eqnarray}
\widehat{\delta}(X)=\frac{1}{2\pi}\int^{\infty}_{-\infty} e^{itX}dt,\qquad
\Gamma(X)=\int^{\infty}_{0} t^{X-1}e^{-t}dt,
\end{eqnarray}
respectively.
The formula (\ref{formulaA6}) can be used to compute the third line in (\ref{A4}), whilst the second lime can be worked out by similar calculations.

\subsection{Calculation of the jumps for the S$^3$ dynamos}
Using the wave solutions, we can explicitly find some key terms \textcolor{black}{on} the left hand side of (\ref{s3roll}). 
\begin{subequations}
\begin{eqnarray}
&&2\int^{\infty}_{-\infty}(|\widetilde{W}_0|^2-|\widetilde{C}_0|^2)dN
\nonumber \\
&=&|\widetilde{Q}_0'|^2\int^{\infty}_{-\infty}\left(\frac{S(\zeta_-)}{(i\alpha |\lambda|)^{2/3}} \right)
\left(\frac{S(\zeta_+)}{(i\alpha |\lambda|)^{2/3}} \right)^*+\left(\frac{S(\zeta_+)}{(i\alpha |\lambda|)^{2/3}} \right)
\left(\frac{S(\zeta_-)}{(i\alpha |\lambda|)^{2/3}} \right)^*dN,\\
&&2\int^{\infty}_{-\infty}(\widetilde{B}_0\widetilde{W}_0^*-\widetilde{V}_0\widetilde{C}_0^*)dN\nonumber \\
&=&-\widetilde{Q}_{00}\widetilde{Q}_0'^*\int^{\infty}_{-\infty}\left (\frac{\kappa(\zeta_-)}{i\alpha |\lambda|}\right )\left (\frac{S(\zeta_+)}{(i\alpha |\lambda|)^{2/3}}\right )^*
-\left (\frac{\kappa(\zeta_+)}{i\alpha |\lambda|}\right )\left (\frac{S(\zeta_-)}{(i\alpha |\lambda|)^{2/3}}\right )^*dN\nonumber\\
&&+\frac{\widetilde{Q}_{00}\widetilde{Q}_0'^*}{2N_0}\int^{\infty}_{-\infty}
\left(\frac{\zeta_-\kappa(\zeta_-)-\zeta_+\kappa(\zeta_+)}{(i\alpha|\lambda|)^{4/3}}\right )
\left(\frac{S(\zeta_-)-S(\zeta_+)}{(i\alpha|\lambda|)^{2/3}}\right )^*dN\nonumber\\
&&-\frac{1}{2N_0}\left (\widetilde{Q}_{00}\widetilde{Q}_0'^*+\frac{2\lambda'|\widetilde{Q}_0'|^2}{3\lambda} \right )\int^{\infty}_{-\infty}\left( \frac{S'(\zeta_-)-S'(\zeta_+)}{(i\alpha |\lambda|)^{4/3}} \right)\left( \frac{S(\zeta_-)-S(\zeta_+)}{(i\alpha |\lambda|)^{2/3}} \right)^*dN\nonumber\\
&&+\left (N_0'+\frac{\lambda'N_0}{3\lambda} \right)2\int^{\infty}_{-\infty}(|\widetilde{W}_0|^2-|\widetilde{C}_0|^2)dN.
\end{eqnarray}
\end{subequations}
In order to derive the final expression of the jumps, the following formulas may be useful.
\begin{subequations}
\begin{eqnarray}
\int^{\infty}_{-\infty}
\left(\frac{\zeta_-\kappa(\zeta_-)-\zeta_+\kappa(\zeta_+)}{(i\alpha|\lambda|)^{4/3}}\right )
\left(\frac{S(\zeta_-)-S(\zeta_+)}{(i\alpha|\lambda|)^{2/3}}\right )^*dN+\text{c.c.}\nonumber \\
=2N_0\int^{\infty}_{-\infty}\left (\frac{\kappa(\zeta_-)}{i\alpha |\lambda|}\right )\left (\frac{S(\zeta_+)}{(i\alpha |\lambda|)^{2/3}}\right )^*
-\left (\frac{\kappa(\zeta_+)}{i\alpha |\lambda|}\right )\left (\frac{S(\zeta_-)}{(i\alpha |\lambda|)^{2/3}}\right )^*dN+\text{c.c.}\nonumber \\
-\int^{\infty}_{-\infty}\left( \frac{S'(\zeta_-)-S'(\zeta_+)}{(i\alpha |\lambda|)^{4/3}} \right)\left( \frac{S(\zeta_-)-S(\zeta_+)}{(i\alpha |\lambda|)^{2/3}} \right)^*dN+\text{c.c.},\\
\int^{\infty}_{-\infty}\left(\frac{S(\zeta_-)}{(i\alpha |\lambda|)^{2/3}} \right)
\left(\frac{S(\zeta_+)}{(i\alpha |\lambda|)^{2/3}} \right)^*dN+\text{c.c.}=\frac{2G_0((\alpha |\lambda|)^{1/3}N_0)}{(\alpha |\lambda|)^{5/3}},\\
\int^{\infty}_{-\infty}\left( \frac{S'(\zeta_-)-S'(\zeta_+)}{(i\alpha |\lambda|)^{4/3}} \right)\left( \frac{S(\zeta_-)-S(\zeta_+)}{(i\alpha |\lambda|)^{2/3}} \right)^*dN=-\frac{2N_0^2G_1((\alpha |\lambda|)^{1/3}N_0)}{(\alpha |\lambda|)^{5/3}}.
\end{eqnarray}
\end{subequations}

\section{Small magnetic Prandtl number cases}

Here, we perform \textcolor{black}{a} similar analysis as \textcolor{black}{in} section 3 but assume $P_m \ll 1$. 
We shall shortly see that the magnetic roll-streak field satisfies a diffusion equation so it must be supported by some constant external magnetic fields. This means that we are concerned with the large Reynolds number asymptotic analysis of so-called inductionless limit.

\subsection{The case $P_m \sim O(R^{-1})$ or larger}

The appropriate leading order asymptotic expansion is
\begin{eqnarray}
\left[ \begin{array}{c} u\\ v\\ w\\a\\b\\c\\q \end{array} \right]=
\left[ \begin{array}{c} \overline{u}\\ R^{-1}\overline{v}\\ R^{-1}\overline{w}\\ \overline{a}\\ R_m^{-1}B+R^{-1}\overline{b}\\ R_m^{-1}C+R^{-1}\overline{c}\\  R^{-2}\overline{q}  \end{array} \right]
+\epsilon R^{-1}\left \{ e^{i \alpha (x-st)}
\left[ \begin{array}{c} \widetilde{u}\\ \widetilde{v}\\ \widetilde{w}\\ \widetilde{a}\\ \widetilde{b}\\ \widetilde{c}\\\widetilde{q}\end{array} \right]
+\text{c.c.}
\right \}.\label{expsmallP1}
\end{eqnarray}
The constants $B,C$ are the magnetic field applied externally.
Using these asymptotic forms in the governing equations and neglecting the small terms, we have the vortex equations
\begin{subequations}
\begin{eqnarray}
~[(\overline{v}\partial_y +\overline{w}\partial_z)-\triangle_2]
\left[ \begin{array}{c} \overline{u}\\ \overline{v}\\ \overline{w} \end{array} \right]
+\left[ \begin{array}{c} 0\\ \overline{q}_y\\ \overline{q}_z \end{array} \right]
=
 \left[ \begin{array}{c} (B\partial_y +C\partial_z)\overline{a}\\ (B\partial_y +C\partial_z)\overline{b}\\ (B\partial_y +C\partial_z)\overline{c} \end{array} \right]\nonumber \\
 -\epsilon^2\left[ \begin{array}{c} 0\\ 
\{ (|\widetilde{v}|^2-|\widetilde{b}|^2 )_y +( \widetilde{v} \widetilde{w}^* -\widetilde{b} \widetilde{c}^*)_z \}+\text{c.c.} \\ 
\{  (|\widetilde{w}|^2-|\widetilde{c}|^2 )_z + ( \widetilde{v} \widetilde{w}^* -\widetilde{b} \widetilde{c}^*)_y \}+\text{c.c.}
\end{array} \right],~~~\label{invmop}
\\
~ -\triangle_2
\left[ \begin{array}{c} \overline{a}\\ \overline{b}\\ \overline{c} \end{array} \right]
-[B\partial_y +C\partial_z]
\left[ \begin{array}{c} \overline{u}\\ \overline{v}\\ \overline{w} \end{array} \right]=
\mathbf{0},\label{diffmag1}
\\
\overline{v}_y+\overline{w}_z=0, \qquad \overline{b}_y+\overline{c}_z=0, ~~~~~
\end{eqnarray}
\end{subequations}
and the wave equations 
\begin{subequations}
\begin{eqnarray}
\left \{
Ui\alpha 
\left[ \begin{array}{c} \widetilde{u}\\ \widetilde{v}\\ \widetilde{w} \end{array} \right] 
+
\left[ \begin{array}{c} \widetilde{v}U_y+\widetilde{w}U_z\\ 0\\ 0 \end{array} \right]
\right \} \hspace{60mm}\nonumber \\
-
\left \{
[\overline{a}i\alpha+R_m^{-1}B\partial_y+R_m^{-1}C\partial_z]
\left[ \begin{array}{c} \widetilde{a}\\ \widetilde{b}\\ \widetilde{c} \end{array} \right] 
+
\left[ \begin{array}{c} \widetilde{b}\overline{a}_y+\widetilde{c}\overline{a}_z\\ 0\\ 0 \end{array} \right]
\right \}
+\left[ \begin{array}{c} i\alpha \widetilde{q} \\ \widetilde{q}_y \\ \widetilde{q}_z \end{array} \right]=R^{-1}\triangle 
\left[ \begin{array}{c} \widetilde{u}\\ \widetilde{v}\\ \widetilde{w} \end{array} \right],~~~\label{Cwavemop}
\\
\left \{
Ui\alpha
\left[ \begin{array}{c} \widetilde{a}\\ \widetilde{b}\\ \widetilde{c} \end{array} \right] 
+
\left[ \begin{array}{c} \widetilde{v}\overline{a}_y+\widetilde{w}\overline{a}_z\\ 0\\ 0 \end{array} \right]
\right \} \hspace{70mm} \nonumber \\
-
\left \{
[\overline{a}i\alpha+R_m^{-1}B\partial_y+R_m^{-1}C\partial_z]
\left[ \begin{array}{c} \widetilde{u}\\ \widetilde{v}\\ \widetilde{w} \end{array} \right] 
+
\left[ \begin{array}{c} \widetilde{b}U_y+\widetilde{c}U_z\\ 0\\ 0 \end{array} \right]
\right \}
=R_m^{-1}\triangle 
\left[ \begin{array}{c} \widetilde{a}\\ \widetilde{b}\\ \widetilde{c} \end{array} \right],~~~\label{Cwaveindp}\\
i\alpha \widetilde{u}+\widetilde{v}_y+\widetilde{w}_z=0,\qquad i\alpha \widetilde{a}+\widetilde{b}_y+\widetilde{c}_z=0.~~~~~~~~~~
\end{eqnarray}
\end{subequations}
Here, by assumption, $R_m^{-1}$ is $O(1)$ or smaller.
If both $B,C$ are zero, (\ref{diffmag1}) is merely a diffusion equation and, thus, there is no non-trivial solution for homogeneous boundary conditions. 

The size of $R_m$ controls the existence of the singularity in the flow. 
When $R_m$ is $O(1)$, there is a diffusion effect in the wave equations and, thus, there is no singularity at all. 
The viscous terms in the momentum equations can safely be switched off everywhere, and we must set $\epsilon=1$ to balance the wave Reynolds stress term in the vortex equations. 
If $R_m^{-1}$ is smaller than $O(1)$, the dissipative effect is absent in the wave equations and thus there is the Alfv\'en resonant point singularity. However the analysis of the dissipative layer is the same as \textcolor{black}{in} section 3, and so, we omit that detail here.

\subsection{The case $P_m\ll R^{-1}$ }

Now $R_m$ is set to be asymptotically small. The governing equations suggest that we can further rescale the magnetic fields by factor of $R_m^{1/2}$ from (\ref{expsmallP1}):
\begin{eqnarray}
\left[ \begin{array}{c} u\\ v\\ w\\a\\b\\c\\q \end{array} \right]=
\left[ \begin{array}{c} \overline{u}\\ R^{-1}\overline{v}\\ R^{-1}\overline{w}\\ \overline{a}\\ R_m^{-1/2}B+R_m^{1/2}R^{-1}\overline{b}\\ R_m^{-1/2}C+R_m^{1/2}R^{-1}\overline{c}\\  R^{-2}\overline{q}  \end{array} \right]
+\epsilon R^{-1}\left \{ e^{i \alpha (x-st)}
\left[ \begin{array}{c} \widetilde{u}\\ \widetilde{v}\\ \widetilde{w}\\ R_m^{1/2}\widetilde{a}\\ R_m^{1/2}\widetilde{b}\\ R_m^{1/2}\widetilde{c}\\\widetilde{q}\end{array} \right]
+\text{c.c.}
\right \},
\end{eqnarray}
Substitution of this asymptotic expression into the governing equations yields the leading order vortex system
\begin{subequations}
\begin{eqnarray}
~[(\overline{v}\partial_y +\overline{w}\partial_z)-\triangle_2]
\left[ \begin{array}{c} \overline{u}\\ \overline{v}\\ \overline{w} \end{array} \right]
+\left[ \begin{array}{c} 0\\ \overline{q}_y\\ \overline{q}_z \end{array} \right]
=
 \nonumber \\
 -\epsilon^2\left[ \begin{array}{c} 0\\ 
\{ (|\widetilde{v}|^2-|\widetilde{b}|^2 )_y +( \widetilde{v} \widetilde{w}^* -\widetilde{b} \widetilde{c}^*)_z \}+\text{c.c.} \\ 
\{  (|\widetilde{w}|^2-|\widetilde{c}|^2 )_z + ( \widetilde{v} \widetilde{w}^* -\widetilde{b} \widetilde{c}^*)_y \}+\text{c.c.}
\end{array} \right],~~~\label{invmop}
\\
~ -\triangle_2
\left[ \begin{array}{c} \overline{a}\\ \overline{b}\\ \overline{c} \end{array} \right]
-[B\partial_y +C\partial_z]
\left[ \begin{array}{c} \overline{u}\\ \overline{v}\\ \overline{w} \end{array} \right]=
\mathbf{0},
\\
\overline{v}_y+\overline{w}_z=0, \qquad \overline{b}_y+\overline{c}_z=0, ~~~~~
\end{eqnarray}
\end{subequations}
and the wave system
\begin{subequations}
\begin{eqnarray}
Ui\alpha 
\left[ \begin{array}{c} \widetilde{u}\\ \widetilde{v}\\ \widetilde{w} \end{array} \right] 
+
\left[ \begin{array}{c} \widetilde{v}U_y+\widetilde{w}U_z\\ 0\\ 0 \end{array} \right]
-
[B\partial_y+C\partial_z]
\left[ \begin{array}{c} \widetilde{a}\\ \widetilde{b}\\ \widetilde{c} \end{array} \right] 
+\left[ \begin{array}{c} i\alpha \widetilde{q} \\ \widetilde{q}_y \\ \widetilde{q}_z \end{array} \right]=\mathbf{0},~~~\label{Cwavemop}
\\
-
[B\partial_y+C\partial_z]
\left[ \begin{array}{c} \widetilde{u}\\ \widetilde{v}\\ \widetilde{w} \end{array} \right] 
=\triangle 
\left[ \begin{array}{c} \widetilde{a}\\ \widetilde{b}\\ \widetilde{c} \end{array} \right],~~~\label{Cwaveindp}\\
i\alpha \widetilde{u}+\widetilde{v}_y+\widetilde{w}_z=0,\qquad i\alpha \widetilde{a}+\widetilde{b}_y+\widetilde{c}_z=0.~~~
\end{eqnarray}
\end{subequations}
There is a diffusivity in the wave equations and, thus, there is no resonant layer singularity. 
The appropriate choice of the wave amplitude is $\epsilon=1$. The form of the magnetic roll-streak equation is unchanged from the previous case so the remark made in section B1 is also applicable here.

\end{document}